\documentclass[traditabstract]{aa} 
\usepackage[toc,page]{appendix}
\usepackage{graphicx}
\usepackage{natbib,twoopt,amssymb,lscape}
\usepackage{lipsum}
\usepackage[breaklinks=true,hidelinks]{hyperref} 
\usepackage{xcolor,ulem} 
\bibpunct{(}{)}{;}{a}{}{,} 
\newcommandtwoopt{\citeads}[3][][]{\href{http://adsabs.harvard.edu/abs/#3}%
{\citealp[#1][#2]{#3}}} 
\newcommandtwoopt{\citepads}[3][][]{\href{http://adsabs.harvard.edu/abs/#3}%
{\citep[#1][#2]{#3}}} 
\newcommandtwoopt{\citetads}[3][][]{\href{http://adsabs.harvard.edu/abs/#3}%
{\citet[#1][#2]{#3}}} 
\newcommandtwoopt{\citeyearads}[3][][]%
{\href{http://adsabs.harvard.edu/abs/#3}{\citeyear[#1][#2]{#3}}}
\usepackage{graphicx}
\usepackage{longtable}
\usepackage{natbib}
\bibpunct{(}{)}{;}{a}{}{,} 
\usepackage{soulutf8}
\usepackage{filecontents}
\usepackage{changes}
\usepackage{txfonts}
\newcommand{\gaia}{{\it Gaia}}

\newcommand{\xmm}{{\it XMM-Newton}}

\newcommand{\swift}{{\it Swift}}
\newcommand{\integral}{\textit{INTEGRAL}}

\newcommand{\tess}{TESS}

\newcommand{\lumcgs}{ergs~s$^{-1}$}

\begin{document}

   \title{Symbiotic stars in X-rays IV:  \xmm, \swift~ and \tess\ observations}


   \author{I. J. Lima,  \inst{1,2,3}
            \and
          G. J. M. Luna\inst{4}
          \and
          K. Mukai \inst{5,6}
          \and
          A. S. Oliveira \inst{7}
          \and
          J. L. Sokoloski \inst{8}
          \and
          F. M. Walter \inst{9}
          \and
          N. Palivanas \inst{7}   
          \and
          N. E. Nuñez  \inst{2}
          \and
          R. R. Souza \inst{3}
          \and
          R. A. N. Araujo \inst{3}   
          }

   \institute{CONICET-Universidad de Buenos Aires, Instituto de Astronom\'{i}a y F\'{i}sica del Espacio (IAFE), Av. Inte. G\"{u}iraldes 2620, C1428ZAA, Buenos Aires, Argentina
             \email{isabellima01@gmail.com}
    \and
    Universidad Nacional de San Juan, Facultad de Ciencias Exactas, Físicas y Naturales.,  Av. Ignacio de la Roza 590 (O), Complejo Universitario "Islas Malvinas", Rivadavia, J5402DCS, San Juan, Argentina
    \and
    Universidade Estadual Paulista ``J\'{u}lio de Mesquita Filho", UNESP, Campus of Guaratinguet\'{a}, Av. Dr. Ariberto Pereira da Cunha, 333 - Pedregulho, Guaratinguet\'{a} - SP, 12516-410, Brazil
    \and     
    CONICET-Universidad Nacional de Hurlingham, Av. Gdor. Vergara 2222, Villa Tesei, Buenos Aires, Argentina
    \and
    CRESST and X-ray Astrophysics Laboratory, NASA Goddard Space Flight Center, Greenbelt, MD 20771, USA
    \and 
    Department of Physics, University of Maryland Baltimore County, 1000 Hilltop Circle, Baltimore MD 21250, USA
    \and
    IP\&D, Universidade do Vale do Para\'\i ba, 12244-000, S\~ao Jos\'e dos Campos, SP, Brazil
    \and
    Columbia Astrophysics Lab. 550 W120th St., 1027 Pupin Hall, MC 5247 Columbia University, New York, NC 10027, USA
    \and
    Department of Physics \& Astronomy, Stony Brook University, Stony Brook, NY 11794-3800, USA
              }


 
  \abstract
   {White dwarf symbiotic binaries are detected in X-rays with luminosities in the range of 10$^{30}$ to 10$^{34}$~\lumcgs. Their X-ray emission arises either from the accretion disk boundary layer, from a region where the winds from both components collide or from nuclear burning on the white dwarf surface. In our continuous effort to identify X-ray emitting symbiotic stars, we studied four systems using observations from the Neil Gehrels \swift~Observatory and \xmm\ satellites in X-rays and from \tess\ in the optical. 
   The X-ray spectra were fit with absorbed optically thin thermal plasma models, either single- or multitemperature with kT~$<$~8~keV for all targets.
   Based on the characteristics of their X-ray spectra, we classified BD~Cam as possible $\beta$-type, V1261~Ori and CD~-27~8661 as $\delta$-type, and confirmed NQ~Gem as $\beta$/$\delta$-type. 
   The $\delta$-type X-ray emission most likely arise in the boundary layer of the accretion disk, while in the case of BD~Cam, its mostly-soft emission originates from shocks, possibly between the red giant and WD/disk winds.
   In general, we have found that the observed X-ray emission is powered by accretion at a low accretion rate of about 10$^{-11}$~M$_{\odot}$~yr$^{-1}$. The low ratio of X-ray to optical luminosities, however indicates that the accretion-disk boundary layer is mostly optically thick and tends to emit in the far or extreme UV. The detection of flickering in optical data provides evidence of the existence of an accretion disk. 
   }

   \keywords{(stars:) binaries: symbiotic - accretion, accretion disks - X-rays: individual: BD~Cam, V1261~Ori, NQ~Gem, CD~-27~8661
               }

   \maketitle
%

\section{Introduction}

Symbiotic stars are interacting binaries consisting of a late-type giant and a compact object, generally a white dwarf (WD) (e.g., \citealt{2009Kenyon}). These systems exhibit blended characteristics of both stellar components and in some cases, an accretion disk and an ionized nebula. Typically, their optical spectra show emission lines originated in the nebula (e.g., H~I, He~II, [OIII] lines), TiO absorption features from the cool giant photosphere, and UV and blue continuum radiated by the hot component (e.g., \citealt{Kenyon_1986}). 

Some symbiotic systems present an abundance anomaly due to $s$-process elements enhancement, known as the {\it barium-star syndrome} (e.g., BD-21~3873, AG~Dra, UKS-Ce1, BD Cam, and V1261~Ori; see \citealt{Smith_1996, Smith_1997}). During the evolution in the Asymptotic Giant Branch (AGB), the progenitor of the WD loses its carbon and heavy elements rich envelope through a strong stellar wind, and part of the material is accreted by the less evolved companion, leading to future red giants which in some cases contain Technetium (Tc) \citep{Jorissen_1996}.

The X-ray emission in symbiotic stars originates from nuclear burning, colliding winds, accretion, or a combination of those (see Table~2 in \citealt{Mukai_2017}). In $\alpha$-type X-ray emission systems, the quasi-steady thermonuclear burning on the surface of the accreting WD is responsible for the super-soft emission (kT~$<$~0.4~keV); the X-ray emission in $\beta$-type (kT~$\leq$~2.4~keV) could be produced by colliding winds from the two stars; 
$\gamma$-type systems are symbiotics containing neutron stars emitting hard X-rays with energies kT~$>$~2.4~keV; in $\delta$-type systems the X-ray emission arises from the accretion disk boundary layer, a region between the accretion disk and the WD, with kT of 5 to 50~keV, depending on the WD mass. Finally, in $\beta$/$\delta$-type, the soft component is originated in a colliding-wind region while the hard flux is produced in the boundary layer (e.g., \citealt{Muerset_1997,Luna_2013}). In these WD symbiotic systems, the thermal X-ray emission is optically thin or blackbody-type; in the symbiotics where the compact object is thought to be a neutron star the X-ray emission is due to Comptonization of X-ray photons.
There are 56 confirmed and classified symbiotic stars with X-ray emission, including extra-galactic sources (see list in Table~\ref{tab:x_ray_list}), being 8~$\alpha$-type, 21~$\beta$-type, 7~$\gamma$-type, 11~$\delta$-type, and 9~$\beta$/$\delta$-type (e.g., \citealt{Luna_2013, nunez_2014, nunez_2016, Merc_2019}). A concatenated list of symbiotic binaries is given in ``The New Online Database of Symbiotic Variables" catalog\footnote{Jaroslav Merc's catalog of symbiotic stars at \url{https://sirrah.troja.mff.cuni.cz/~merc/nodsv/}.}(see \citealt{Merc_2019RNAAS...3...28M}).

The symbiotic systems with the longest orbital periods contain WDs, with periods ranging from hundreds of days to a few years, and semi-major axes of a few to tens of AUs, and accretion rates of 10$^{-11}$~--~10$^{-7}$~M$_{\odot}$~yr$^{-1}$, on average greater than those found in cataclysmic variables (CVs). 
The accretion can occur via Roche-lobe overflow or via Bondi~–~Hoyle capture of the red giant's wind (e.g., \citealt{Bondi_1944, Iben_1996}). Some symbiotics with Mira secondary stars shows evidence of wind Roche-lobe overflow mass transfer, which is a combination of standard Roche-lobe overflow and Bondi–Hoyle accretion (e.g., \citealt{Mohamed_2007, Sokoloski_2010}). 

In CVs, stochastic broad-band variability on time scales of minutes to hours (also referred to as flickering) is commonly observed in the optical band and with increasing amplitude toward the ultraviolet (UV). The origin of flickering is not well understood, but several scenarios have been proposed in order to explain it, including unstable mass accretion onto the WD leading to flickering on a disk hotspot, a turbulent accretion disk, magnetic discharges in an accretion disk, and boundary layer instabilities leading to unsteady accretion \citep{Bruch_1992}.
Optical flickering have been detected in symbiotics with hard X-ray emission and highly absorbed components whereas in other symbiotic stars, where the radiation from the red giant and the surrounding wind nebula overwhelms radiation from the disk (e.g., \citealt{Sokoloski_2001, Luna_2007, Luna_2013}), the detection of flickering has been elusive so far. 
The symbiotic systems where flickering with amplitudes greater than 0.1~mag has been detected are: V694~Mon, Z~And, V2116~Oph, CH~Cyg, RT~Cru, $o$~Cet, V407~Cyg, V648~Car, EF~Aql, ZZ~CMi, and SU~Lyn (e.g., \citealt{Zamanov_2021, Zamanov_2023b}). V1044~Cen, ASAS~J190559-2109.4, ASAS~J152058-4519.7, Gaia~DR2~5917238398632196736, and Gaia~DR2~6043925532812301184 were detected by \citealt{2021PhDT........17L} using B band time series from the SkyMapper Southern Sky Survey. Recently, \cite{Merc2023} reported 20 more systems with flickering using \tess\ data, in 13 of them, flickering was  detected for first time. 
The presence of flickering and its amplitude can be seen as an indicator of the presence of the accretion disk.

The study of symbiotic stars is of special significance as they might be candidates for SN Ia progenitors within the single degenerate scenario \citep{Kenyon_1993}.
Also, those systems allow the study of stellar evolution, the mass transfer accretion processes, the stellar winds jets, the dust formation, and the thermonuclear outbursts (e.g., \citealt{Sokoloski_2003}).
In this paper, we analyze the X-ray emission and fast optical photometry of a sample of four symbiotics, two of them, BD~Cam and V1261~Ori, with the barium-syndrome. The paper is organized as follows: Section~\ref{sec:obs} presents the X-ray and optical observations and their data reduction description. Section~\ref{sec:bayes} discuss the X-ray analysis. Section~\ref{sec:timeseries} summarizes the search for flickering in optical data. The results of the optical and X-ray data for each target are described in Section~\ref{sec:res}. Finally, we present a discussion and conclusions in Sections~\ref{discussion} and~\ref{sec:con}, respectively.

\begin{figure*}
\centering
(a) BD Cam\\ \includegraphics[width=14.1cm] {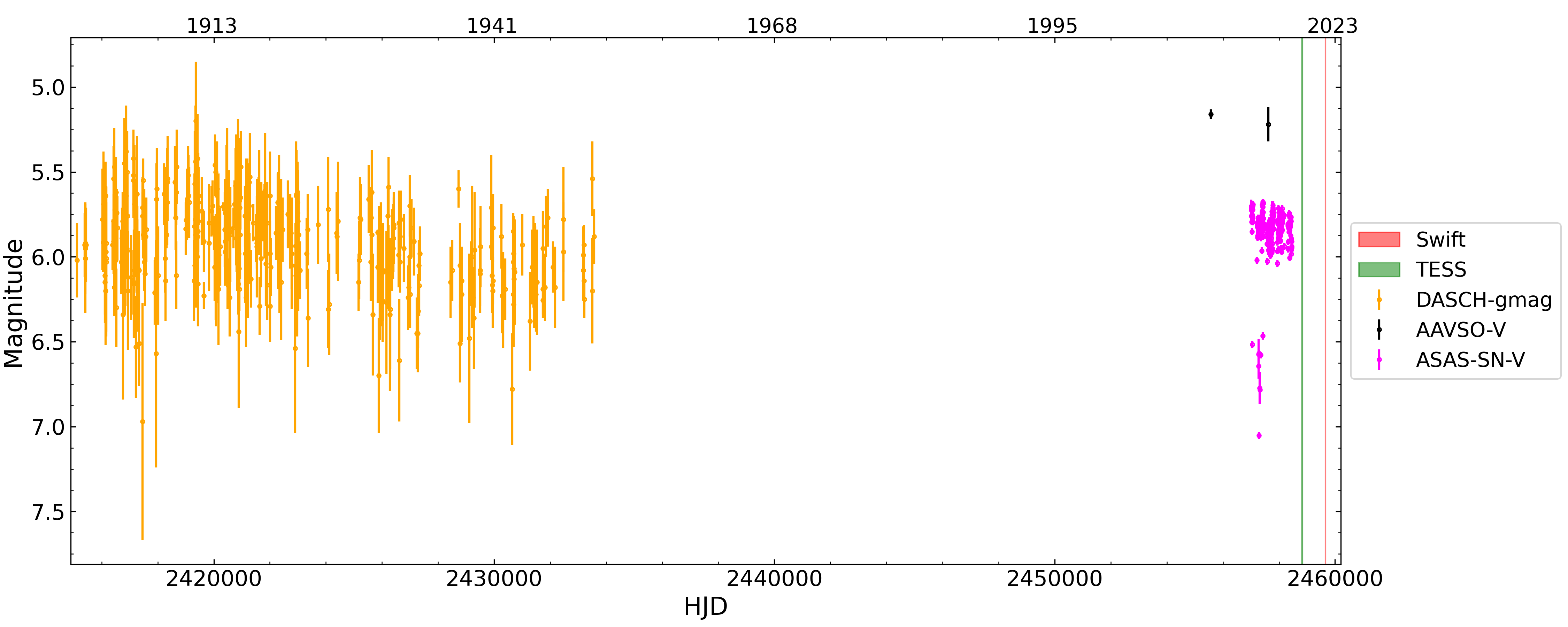}\\
(b) V1261 Ori\\ \includegraphics[width=14.1cm]{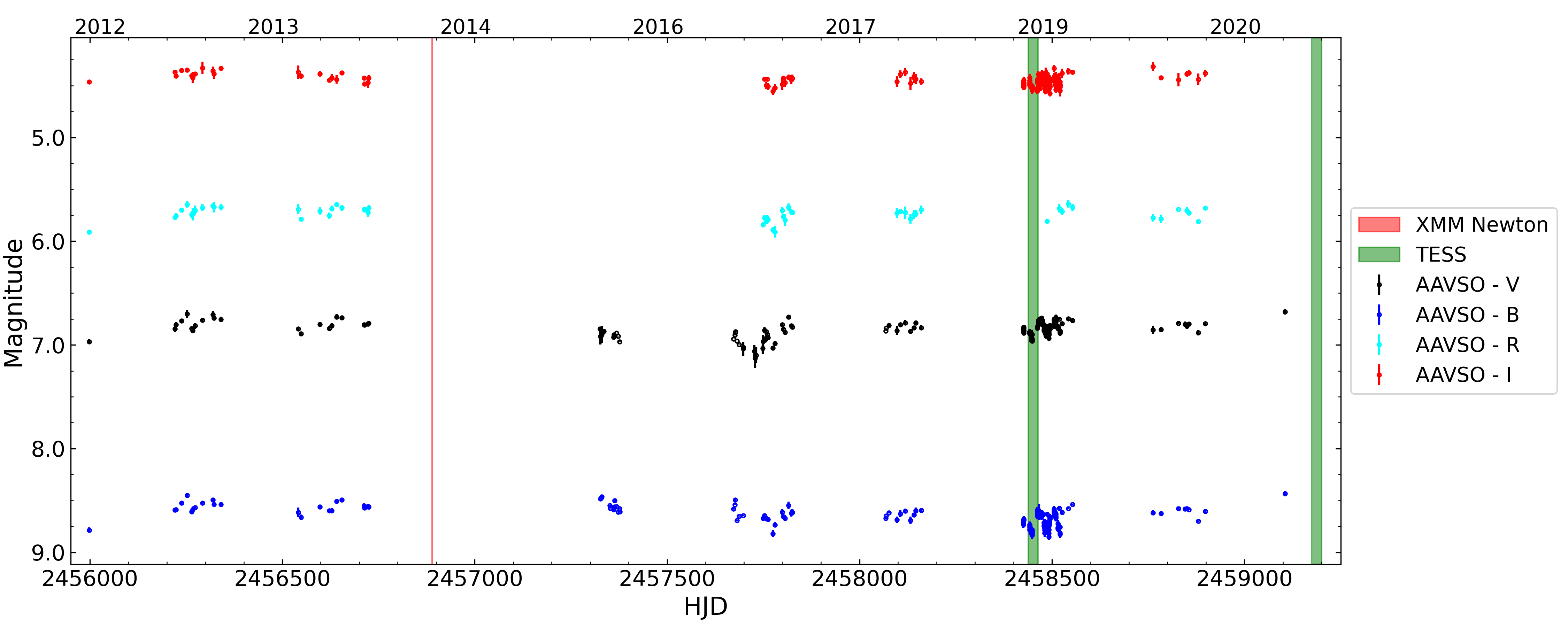}\\
(c) NQ Gem\\ \includegraphics[width=14.1cm]{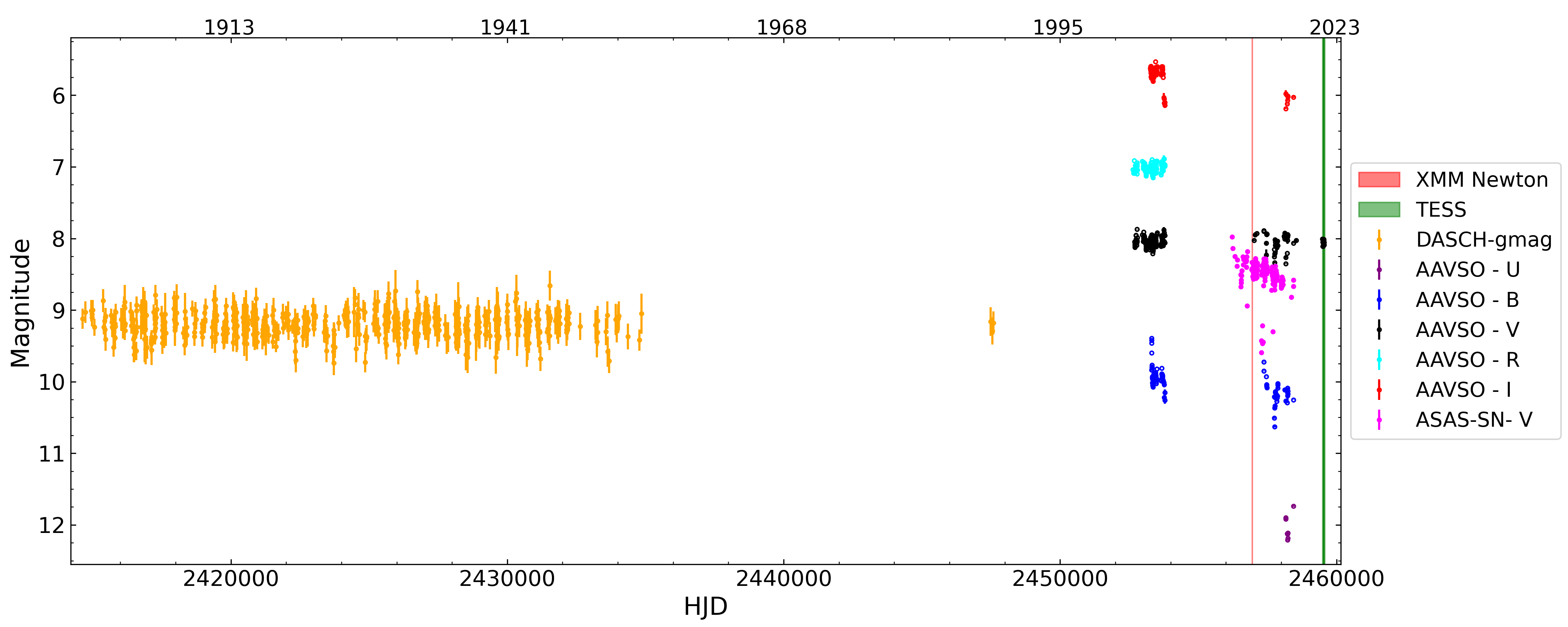}\\ 
(d) CD -27~8661\\ \includegraphics[width=14.1cm]{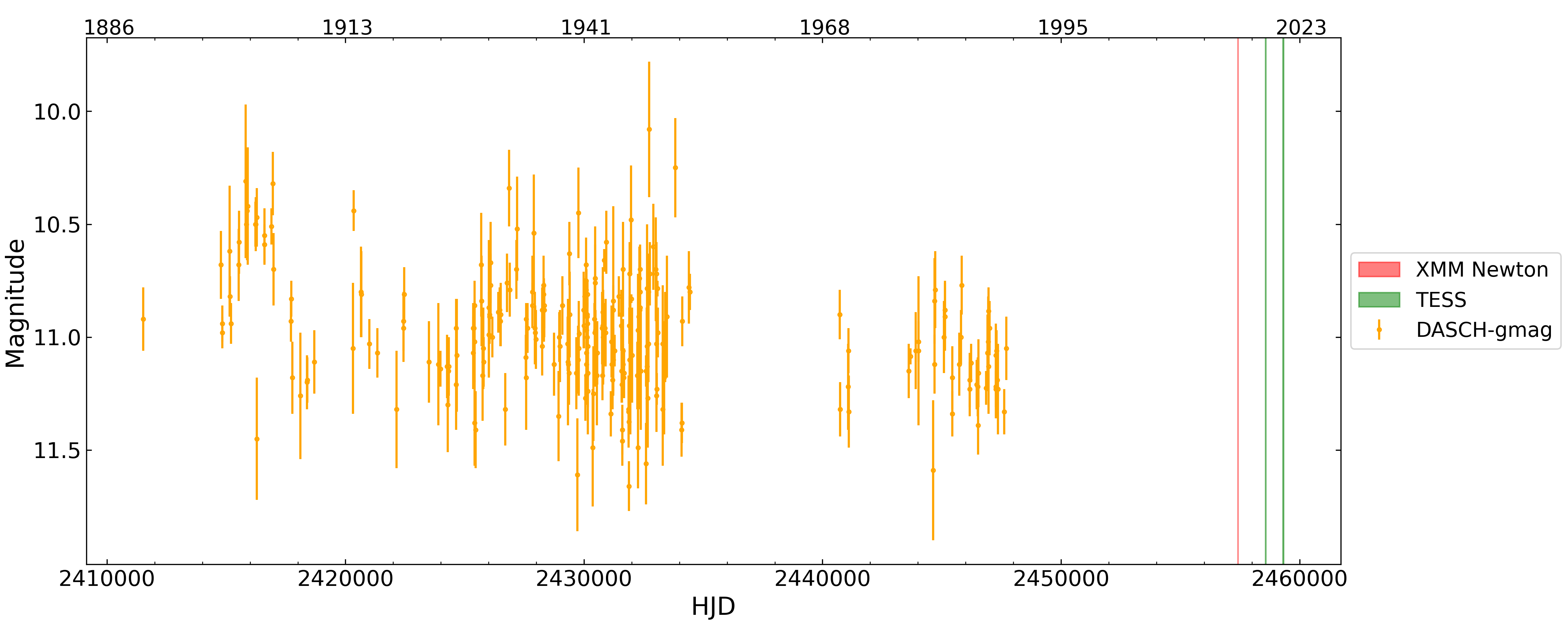} \\ 
\caption{Historical optical photometric observations from {\sc AAVSO}, {\sc ASAS-SN}, {\sc DASCH} of symbiotic systems: (a) BD~Cam, (b) V1261~Ori, (c) NQ~Gem, and (d) CD~-27~8661. The {\sc AAVSO}-UBVRI data are on the Vega system and {\sc ASAS-SN}- V and {\sc DASCH}-gmag data are on the AB system. The red and green lines indicate when the X-ray and \tess\ observations, respectively, were made. Our search did not provide optical measurements during the times that BD~Cam and CD~-27~8661 were observed in X-rays and with \tess\, we thus assume that all observations analyzed here were taken during normal brightness states of the sources.}
\label{fig:historical-light-curves}
\end{figure*}


\section{Observations and data reductions} \label{sec:obs}

We present observations of our symbiotics BD~Cam, V1261~Ori, NQ~Gem, and CD~-27~8661 obtained with \xmm, \swift, and \tess. Details of each observation analyzed here are presented in Table~\ref{tab:log-observations}, where objects are sorted by their right ascension.
In Figure~\ref{fig:historical-light-curves}, we show the times of the \xmm, \swift, and \tess\ observations within the historical optical light curves from the American Association of Variable Star Observers ({\sc AAVSO}\footnote{See AAVSOs Light Curve Generator in
\url{https://www.aavso.org/LCGv2/}.}), the All-Sky Automated
Survey for Supernovae \citep[{\sc ASAS-SN};][]{Shappee_2014, Kochanek_2017}, and the Harvard Digital Access to a Sky-Century at Harvard project ({\sc DASCH}\footnote{See {\sc DASCH} in \url{http://dasch.rc.fas.harvard.edu/lightcurve.php}}; \citealt{Grindlay_2009}). The observations with \xmm, \swift, and \tess\ were obtained during normal flux states for all targets.

We used the \textit{PyXspec}, a \textit{Python} interface to the \textit{XSPEC} version 12.12.0 spectral-fitting software \citep{Arnaud_2016, Gordon_2021} for all X-ray spectral analyzes. We also utilized Bayesian statistical methods in order to improve the parameters estimation of our best-fit models (see Section~\ref{sec:bayes}). 

\begin{table*}
\caption[]{Log of observations. \tess\ sectors are typically observed during 25 days. The cadence is related to full frame images.} 
\centering
\label{tab:log-observations}
\begin{tabular}{cccccc} 
\hline\hline  
\noalign{\smallskip}
Objects  & Satellites & Date & ObsId & Exp. Time & Cadence  \\
& & & & (ks) & (min)\\
\noalign{\smallskip}
\hline
\noalign{\smallskip}   
BD Cam & Swift  & 2022~Mar.~17 & 00015075001 & 2.1 & $-$\\
\\
& \tess\ &  2019~Nov.~28  & tess-s0019-2-2 & $-$ & 30\\
&      &  2022~Nov.~26  & tess-s0059-2-2 & $-$ & 3\\
\hline \noalign{\smallskip}
           & \xmm\ MOS & 2014~Aug.~19 &  & 58& $-$\\
& \tess\ & 2018~Nov.~15 & tess-s0005-2-3 & $-$ & 30\\
&      & 2020~Nov.~20 & tess-s0032-2-3 & $-$ & 10\\  
\hline \noalign{\smallskip} 
NQ Gem & \xmm\ pn & 2014~Oct.~26 & 00675230101 & 42 &  $-$ \\
       & \xmm\ MOS & 2014~Oct.~26 & 59 &  $-$ & $-$\\
\\
& \tess\ & 2021~Oct.~12   & tess-s0046-1-3 &  $-$ & 10\\ 
&      & 2021~Nov.~07   & tess-s0045-2-3 &  $-$ & 10\\
&      & 2021~Dec.~02   & tess-s0044-3-4 &  $-$ & 10\\
&      & 2023~Oct.~16   & tess-s0071-4-1 &  $-$ & 0.3\\ 
&      & 2023~Nov.~11   & tess-s0072-2-3 &  $-$ & 0.3\\ 
\hline \noalign{\smallskip}
CD -27 8661 & \xmm\ & 2016~Jan.~27 & 00761720101 & 30 &  $-$\\ 
\\
& \tess\ & 2019~Mar.~26 & tess-s0010-1-1 &  $-$ & 10\\
&      & 2021~Apr.~02 & tess-s0037-1-1 &  $-$ & 30\\
&      & 2023~Apr.~06 & tess-s0064-1-1 &  $-$ & 3\\
\hline \noalign{\smallskip}
         \end{tabular} 
\end{table*}


\subsection{\xmm} \label{subsec:xmm}

\xmm\ observed three symbiotic stars of our sample: V1261~Ori, NQ~Gem, and CD -27~8661, with the European Photon Imaging Camera (EPIC) pn and MOS instruments as well as with the Reflection Grating Spectrograph (RGS) and Optical Monitor (OM) \citep{Struder_2001, Turner_2001, den_Herder_2001, Mason_2001}, see Table~\ref{tab:log-observations}. The EPIC cameras were operated in full frame mode with the thick filter, except for CD~-27~8661 which was observed with the medium filter.

All \xmm\ data were reduced using SAS~(v20.0.0); the event files were obtained with \texttt{emproc} and \texttt{epproc} tools. The event files were filtered for flaring particle background by retaining intervals when the count rates were $\leq$0.35~c~s$^{-1}$ for EPIC-MOS and $\leq$0.4~c~s$^{-1}$ for EPIC-pn\footnote{See SAS threads website in \url{https://www.cosmos.esa.int/web/xmm-newton/sas-threads}.}. 
The spectra and light curves were extracted from circular regions centered on the {\it SIMBAD} coordinates of our targets with radii of 20$^{"}$ (400 pixels).
The background spectra were extracted from a source-free region on the same CCD where our targets were detected. The response matrices were created using the \texttt{arfgen} and \texttt{rmfgen} tools. 

The number of counts detected with the RGS were insufficient to allow us to include them in the spectral or timing analysis. 
We also obtained simultaneous optical fast photometry from OM using the UVW1 ($\lambda_{max}$~=~2675$\AA$) and the UVM2($\lambda_{max}$~=~2205$\AA$) filters for all objects observed by \xmm. 
These observations with the \xmm\ OM were not affected by the Jupiter Patch as they were taken earlier.
We extracted the light curves with 120~s bin size using the SAS tool \texttt{omfchain}.

\begin{figure}[h]
\centering
\includegraphics[width=\hsize]{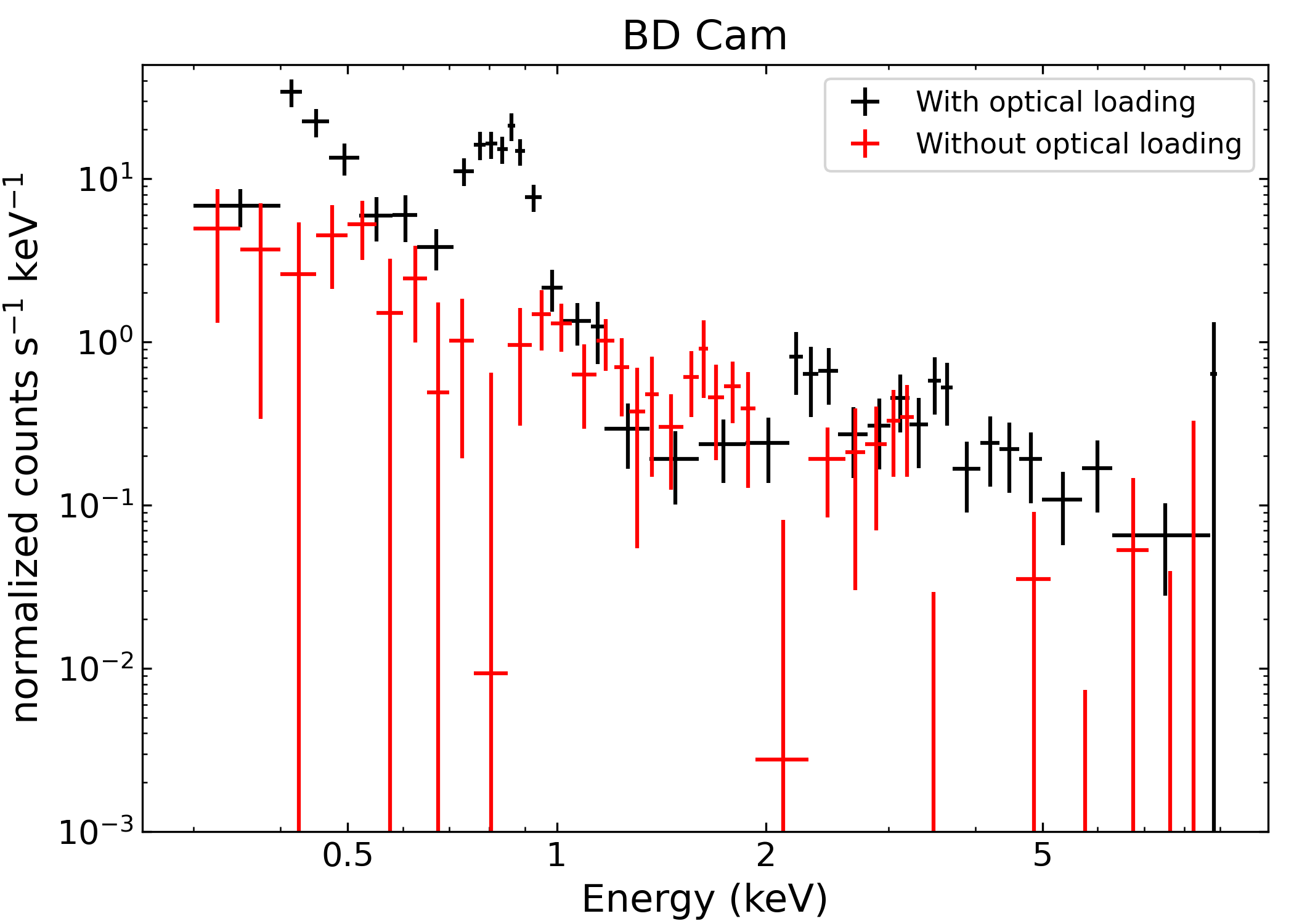}\\
\vspace{0.5cm}
\includegraphics[width=\hsize]{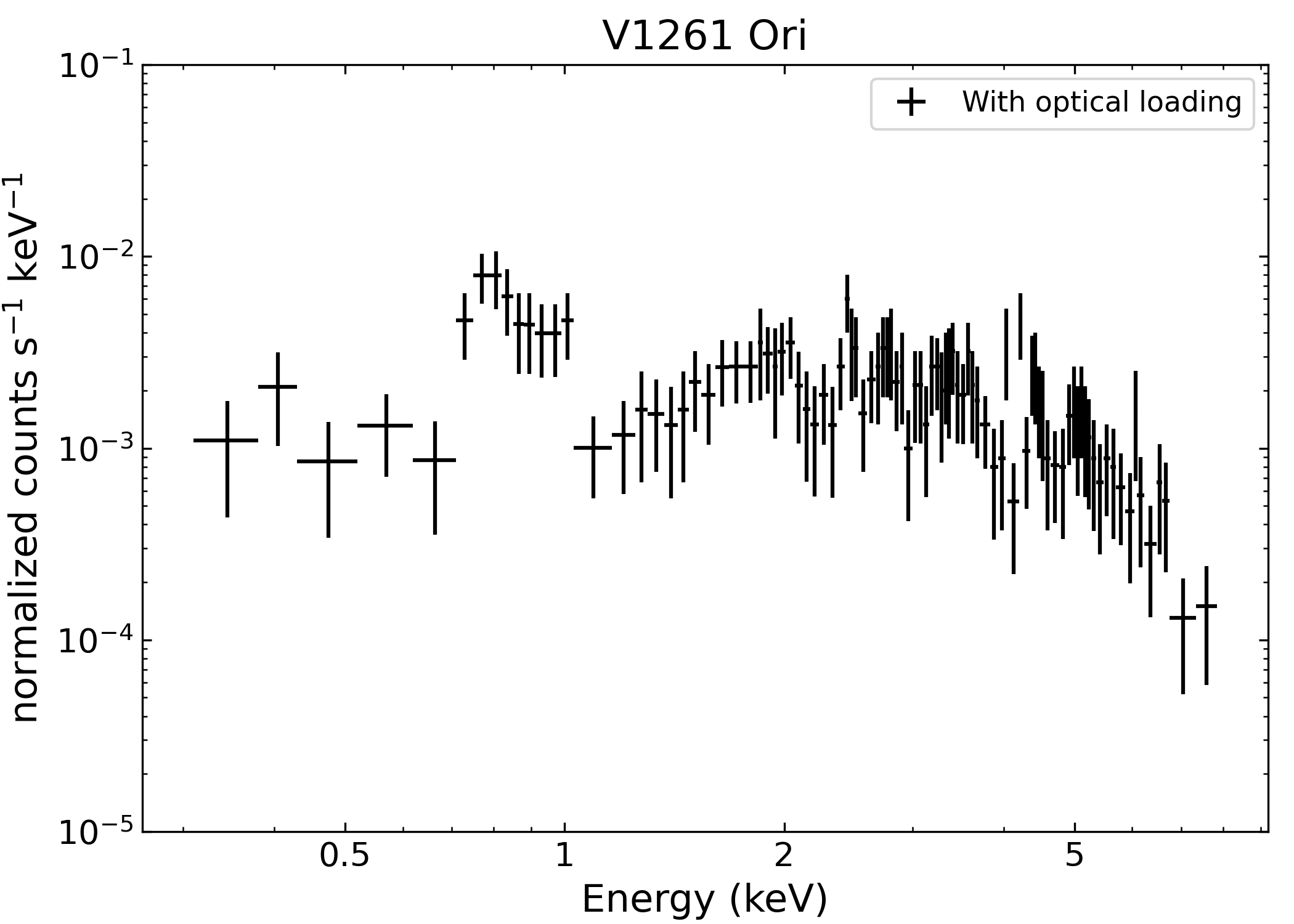}
\caption{Top panel: \swift\ spectra of BD~Cam, strongly affected by optical loading in XRT-PC mode (black) and without (or negligible) effect of optical loading in XRT-WT mode (red). Bottom panel: spectrum of V1261~0ri showing optical loading from \swift~in XRT-PC mode. All data are prepared by the online \swift-XRT data products generator.}
\label{optical_loading}
\end{figure}

\subsection{\swift} \label{subsec:swift}

The \swift\ X-ray Telescope (XRT) observes a sample of  symbiotic stars as part of a survey to search for X-ray emission under multicycle-approved fill-in programs started in \cite{Luna_2013}. The observations were usually performed during approximately $\leq$10~ks (not continuously observed) in XRT using photon counting (PC) mode. 
An additional observation of BD~Cam was recently obtained as a target of opportunity (ToO) in windowed timing (WT) mode in order to minimize the effects of optical loading \citep{Lima_2022}. 
Optical loading distorts the spatial and spectral distributions of the photons producing spurious events in the CCD due to optical photons on-board originated from a bright source.
This effect does not depend on the exposure time of objects, but on the readout time of the detector. \swift/XRT-WT mode, with readout time $\sim$1500 times shorter than \swift/XRT-PC mode, has less probability of optical loading. The strength of optical loading also depends on the spectral type, being more intense for red sources\footnote{See more information about optical loading in \url{https://www.swift.ac.uk/analysis/xrt/optical\_loading.php\#holes}.}.

Some simple tests can be done in order to identify optical loading in X-ray data. First, the image extracted from the event file can present a hole in the center of the source due to rejection in the center of the PSF, in other words, the count-rates are underestimated. 
Second, X-ray photons are scattered to other grades around the source and spurious photons are recorded on the spectra.
Third, we also can assume that the source exhibits the same spectral model for the \swift\ observations as well as \xmm, with only difference in the normalization levels, therefore using the cross-calibration of both observatories, provided by International Astronomical Consortium for High-Energy Calibration (IACHEC)\footnote{See \url{https://iachec.org/}.}, we can measure the count rate.

BD~Cam and V1261~Ori were observed with \swift-XRT in PC mode during three pointings in 2012 April 13, 16, and 17 and fourteen pointings distributed between 2001 April 1 to 2018 May 3, respectively. These objects are very bright in optical wavelengths (V~=~5.11~mag, J~=~1.31~mag for BD~Cam and V~=~8.57~mag, J~=~3.09~mag for V1261~Ori) hence those spectral data are significantly distorted by optical loading and should not be considered for scientific analysis, casting doubts on the results presented by \cite{Merc_2019}.

The impact of optical loading in BD~Cam and V1261~Ori spectra is evident in Figure~\ref{optical_loading}, with sharp features in the region where thermal plasma emits lines of Fe~XVI, Fe~XX, O~VII, O~VIII, etc\footnote{See \url{http://www.atomdb.org/Webguide/webguide.php}.}. Typically, optical loading photons will dominate below $\sim$0.4~keV and will be dramatically reduced above 2~keV \citep{Hare_2021}. However, the X-ray photons at higher energies are not likely real, because they still suffer from grade migration, which induces an overall gain shift and loss of flux, implying that optical loading can affect the entire spectrum. 

For those \swift\ observations that were not affected by optical loading, we obtained the X-ray spectra from \swift\ using the online \swift-XRT data products generator\footnote{See build \swift-XRT products in \url{https://www.swift.ac.uk/user_objects/}.} provided by the UK \swift\ Science Data Centre at the University of Leicester \citep{Evans_2009}. We extracted the X-ray spectra from the event files in 0~--~12 grades using the \texttt{xselect} and identified the centroid of the source by \texttt{xrtcentroid} from a circular radius of 20 pixels ($\approx$47$^{"}$). We used the \swift\ CALDB version 20211108. The background was extracted from an annular region with inner and outer radius of 50 and 60 pixels, respectively. For spectra, we extracted the response ancillary matrix with \texttt{xrtmkarf} tool and the response matrix was provided by \swift\ calibration team. 
In addition, the images from UVOT mode were also obtained 
in UVW1 (2600\AA), UVM2 (2246\AA) and UVW2 (1928\AA) filters.

\subsection{\tess} \label{subsec:tess}

The Transiting Exoplanet Survey Satellite (\tess) covers a broad bandpass (600~--~1000~nm) \citep{Ricker_2015}. \tess\ observations are divided in sectors covering a strip in the sky of 24~$\times$~90 degrees during 27~d and two spacecraft orbits with short gaps for data transmission to Earth.
One of our targets, NQ~Gem, does not have a pipepline-processed light curve in the archive, we thus extracted its light curve, as well as that of the other targets, by first downloading the \tess\ Full Frame Images (FFIs), and then extracted the light curves, using the Python package \texttt{lightkurve} \citep{Lightkurve_2018}.
For each FFI, we generated 30~$\times$~30 pixel Target Pixel File (TPF) cutouts centered on the designated target. Subsequently, the light curves were derived from these cutouts through the quantification of the flux within aperture masks encompassing the target, followed by the subtraction of the average sky flux acquired from a background mask. The identification of our target and other adjacent sources was performed using the \gaia\ catalog. 
BD~Cam was saturated in the FFIs, therefore its \tess\ data were not analyzed in this work (see Figure~2 in \citealt{Lima_2023BAAA...64...59L}).
\section{X-ray analysis} \label{sec:bayes}  

We used the maximum likelihood C-statistic (\textit{cstat}, \citealt{Cash_1979}) when modeling the spectra, which were grouped at one count per bin. C-statistic is based on the Poisson likelihood which is in agreement with the Poisson nature of X-ray data. The $\chi^{2}$ statistic is not appropriate for our sources given their low numbers of counts. In addition, the Bayesian statistic is a powerful approach to compare between models and estimate uncertainties of the model parameters \citep{van_Dyk_2001}. 
We thus have used the resulting parameters from the spectral modeling using the C-statistics as priors for the Bayesian analysis. 

We used the Bayesian X-ray Analysis (BXA) algorithm which uses the \texttt{UltraNest} and fitting environment \textit{XSPEC} \citep{Buchner_2014}. The \texttt{UltraNest}\footnote{\url{https://johannesbuchner.github.io/UltraNest/}.} package derives the posterior probability distributions and the Bayesian
evidence with Monte Carlo algorithm \texttt{MLFriends} \citep{Buchner_2021}. 
We used the log-uniform prior distribution for scale variables such as normalization and $N{_H}$, and uniform prior distribution over other free parameters. Table~\ref{bayes_analysis_priors} lists the priors for each one of the free parameters used to fit the X-ray data using Bayes approach. 
An alternative method of analyzing the quality of a model is the quantile-quantile ($Q-Q$) plots, that integrate observed counts against expected counts from the model (see Appendix A in \citealt{Buchner_2021}). The good fit should be a straight line (slope = 1). The goodness-of-fit was calculated by the Kolmogorov-Smirnov ($K-S$) statistic test, $K~-~S$~=~$sup_{E}\left|observed(< E)~-~predicted (<E)\right|$, which quantified the difference between the cumulative distribution function and the empirical function of observed counts for each energy (E), the value close to 0 provides a better fit. 

\section{Variability}\label{sec:timeseries}

Flickering-like variability on time scales of minutes-to-hours of 20 symbiotics was recently reported by 
\cite{Merc2023} from \tess. From those sources, 13 new systems with no previous detection of flickering were found, including NQ Gem and CD -27 8661 (see their Table~2).
The presence of flickering has erratic pattern in 3 sources similar to symbiotic star RT~Cru. In 59 systems no flickering was detected, neither in \tess\ nor in previous literature (see Table~A.1 of \citealt{Merc2023}). 

We also searched for short-term variability in the exquisite \tess\ and \xmm-OM light curves of our targets. We have first removed the long-term (days) variability from each portion (before and after downlink intervals) of each \tess\ sector and \xmm-OM data by subtracting a Savitzky-Golay (SG; \citealt{Savitzky_1964}) filter. We quantify its flickering strength by the ratio between the observed rms amplitude and that expected from Poisson statistics (see e.g., \citealt{Sokoloski_2001, Luna_2013}). The percentage of fractional rms variability amplitude ($s_{frac}$) can given by Equation~\ref{amplitude}:

\begin{equation}
    s_{frac} = \frac{s} {<flux>} \times 100,
\label{amplitude}
\end{equation}

\noindent where $s$ is the measured standard deviation (in the case of $s/s_{exp}\approx$1, we used $s_{exp}$ and thus $s_{frac}$ is considered an upper limit), $<flux>$ is the mean flux after subtracting the long term trend, and $s_{exp}$ is the expected standard deviation from Poisson statistics. 

The higher ratio of $s$ to $s_{exp}$ indicates that the variation is intrinsic to the source. We also used a field star from \tess\ data in order to give us some confidence that flickering is not an artifact.

\section{Results} \label{sec:res}

In this section, we discuss the results obtained by the analysis of the observations from \xmm, \swift, and \tess.

\begin{figure*}[!htb]
\centering
\includegraphics[scale=0.42]{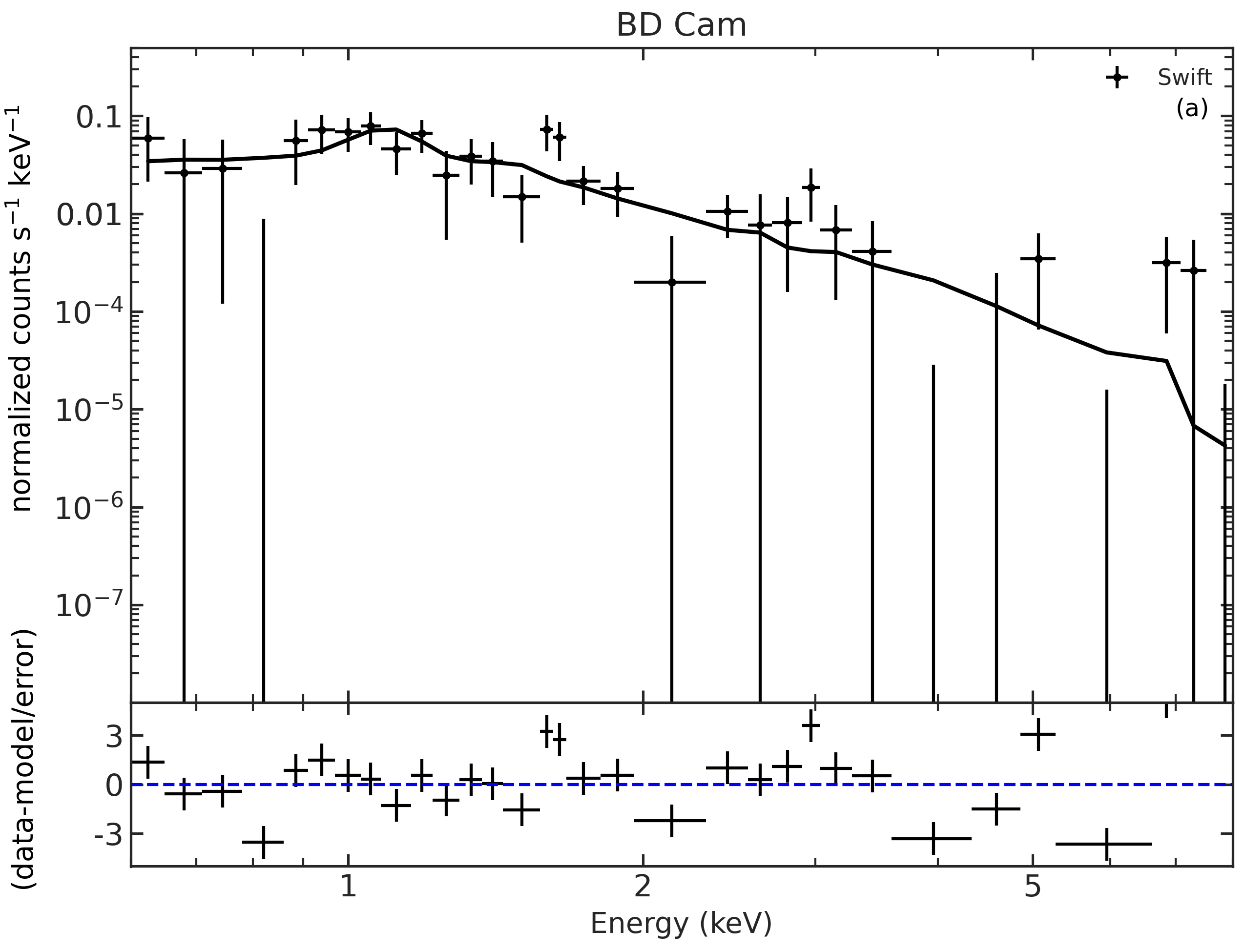}
\includegraphics[scale=0.42]{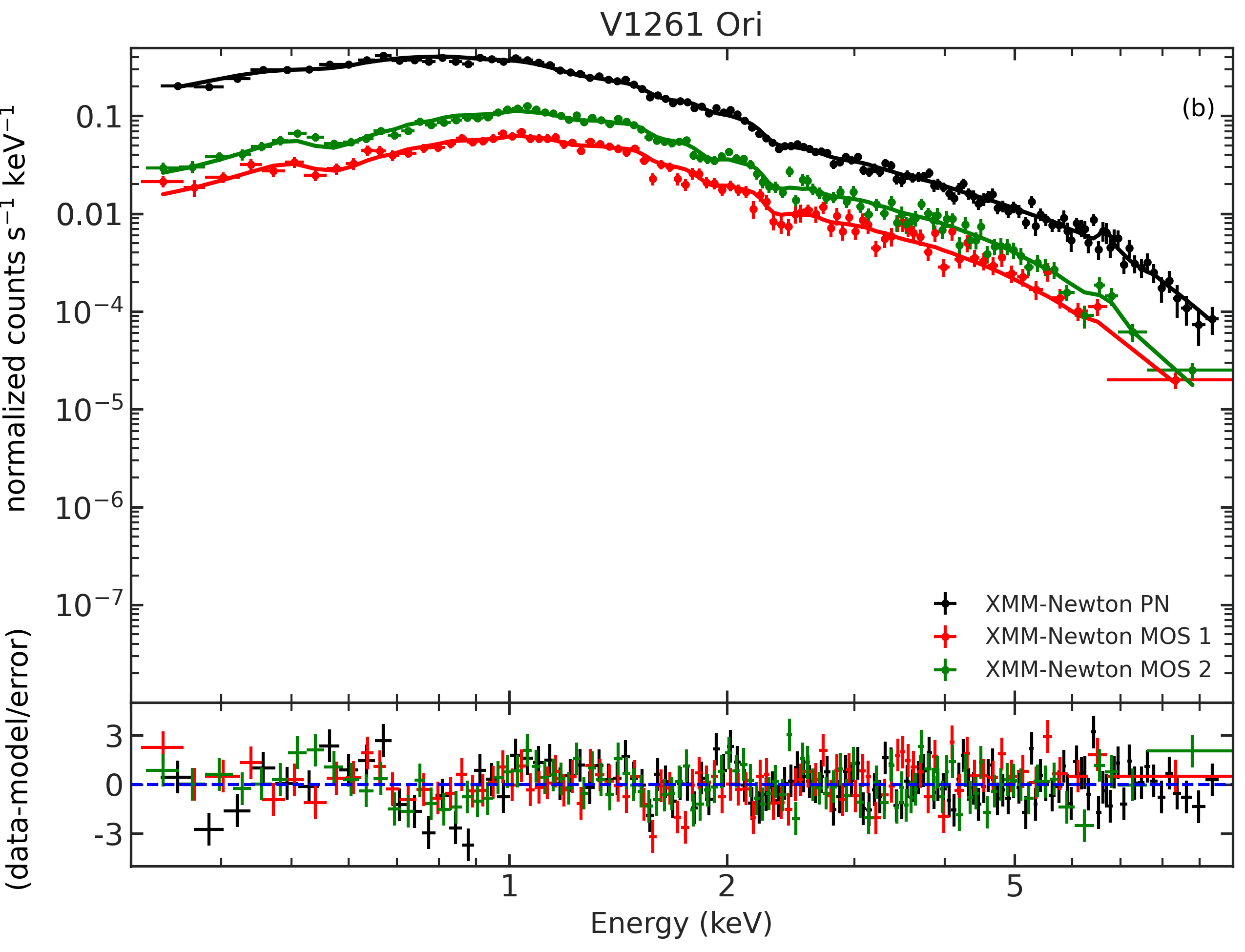}
\includegraphics[scale=0.42]{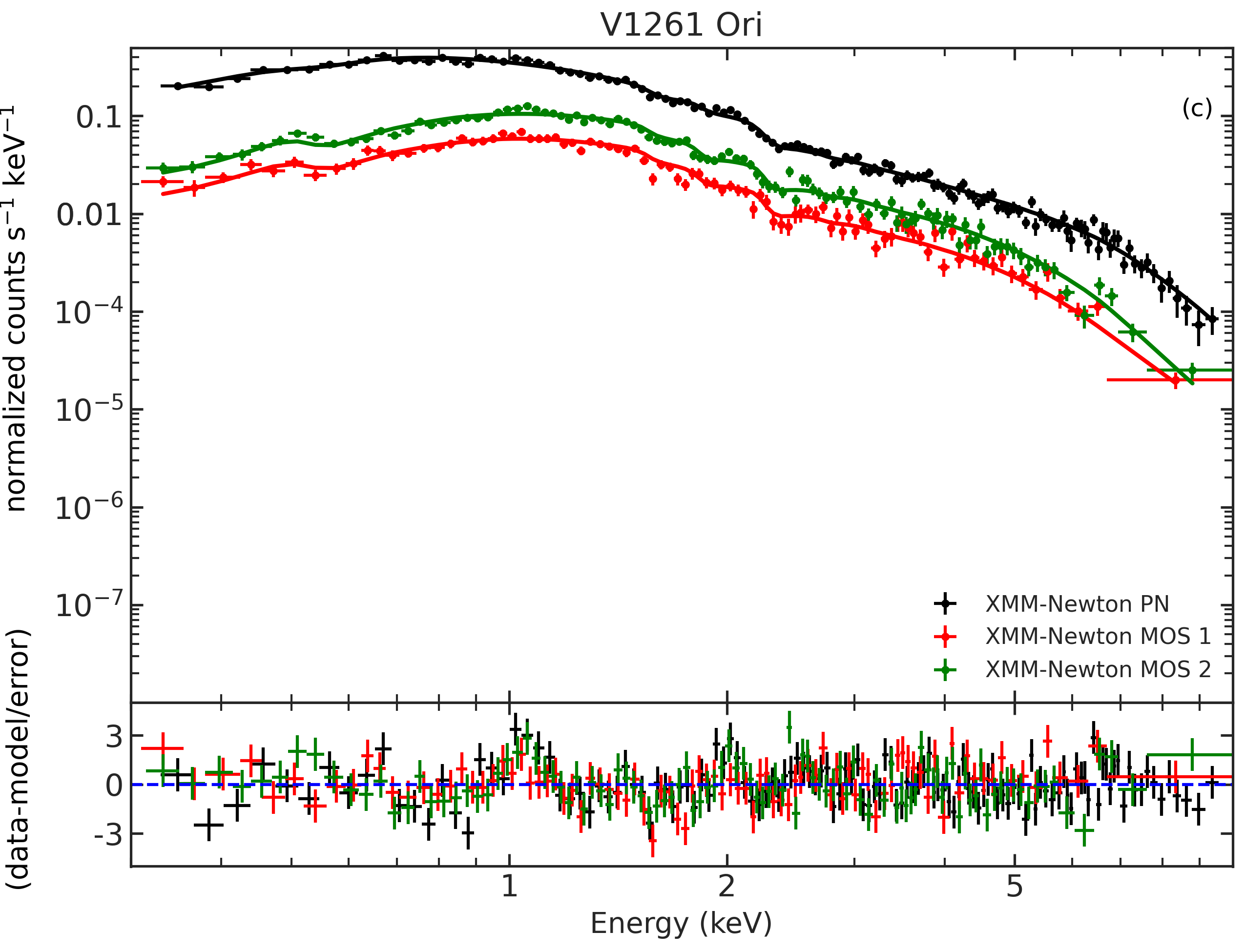}
\includegraphics[scale=0.42]{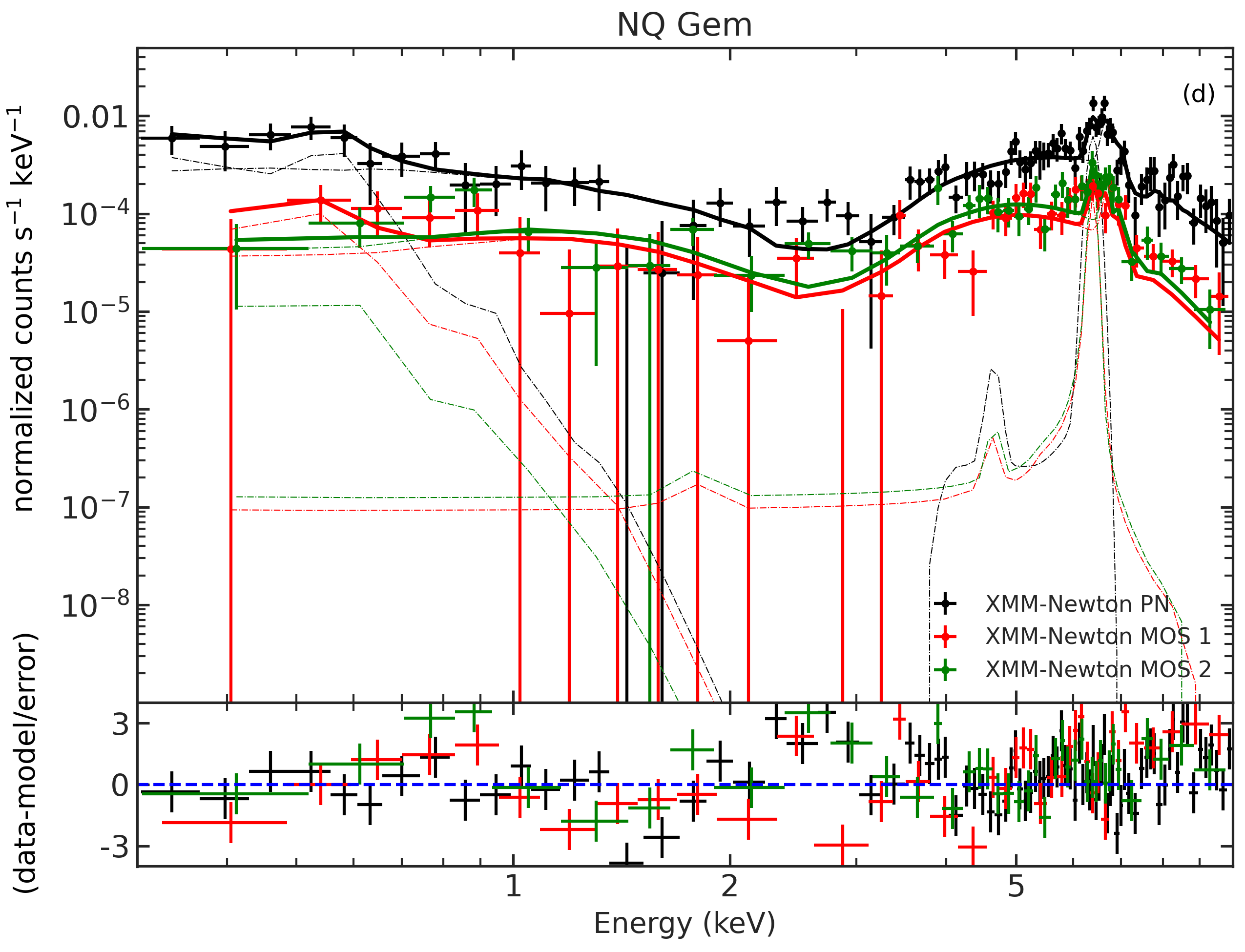}
\includegraphics[scale=0.42]{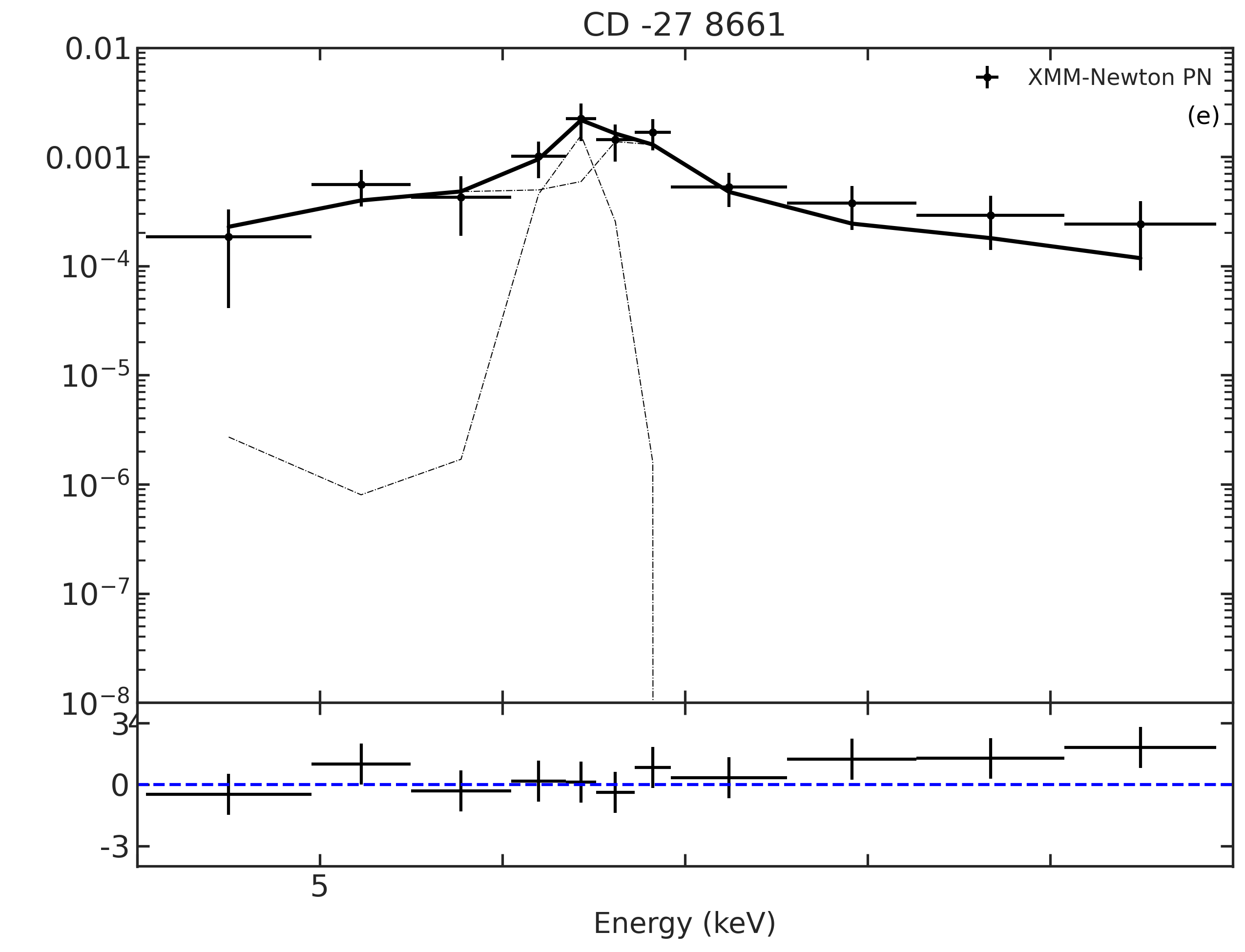}
\caption{X-ray spectra of the symbiotic stars: (a) BD~Cam using \swift\ data, (b) V1261~Ori using \texttt{tbabs~$\times$~mkcflow} model and (c) V1261~Ori using \texttt{tbabs~$\times$~compTT} model, (d) NQ~Gem, and (e) CD~-27~8661. These data were obtained with \xmm\ where the black color is EPIC-pn, the red is EPIC-MOS~1, and the green is EPIC-MOS~2. The solid lines show the best-ﬁt models given in Table~\ref{tab:spectral-models}. The dotted lines show the contribution of the individual spectral components in the case of multicomponent models.}
\label{fig:spectra-analises}
\end{figure*}
\begin{sidewaystable*}
\caption[]{Spectral Models.}
\centering
\small
\label{tab:spectral-models}
\begin{tabular}{ccccccccc} 
\hline\hline  
\noalign{\smallskip}
Objects  & Model & Count Rate & N$_{H}$  &  kT$^{a}$ & Abund & F$_{X}$ & L$_{X}$ & $\chi^{2}$ /dof  \\
         &       & (counts/s) & (10$^{22}$~cm$^{-2}$) & (keV) & &  (10$^{-13}$~erg~s~$^{-1}$~cm$^{-2}$) &  (10$^{31}$ erg~s~$^{-1}$)\\
\noalign{\smallskip}
\hline
\noalign{\smallskip}
BD Cam & \it{tbabs~$\times$~apec} &  0.154~$\pm$~0.015 & $<$0.5 & 2.3$^{+1.8}_{-0.5}$ & 1.0 & 18$^{+7}_{-1}$ & 1.2$^{+0.5}_{-0.2}$ & 2096/154 \\
& & & & & & & \\
\hline \noalign{\smallskip}
V1261 Ori & \it{tbabs~$\times$~mkcflow} & 0.574~$\pm$~0.003 &  0.07~$\pm$~0.01 & 7.7~$\pm$~0.3 & 0.17~$\pm$~0.04 & 23.2~$\pm$~0.2 & 3.9~$\pm$~0.3 & 426/324\\  
         & \it{tbabs~$\times$~compTT} &  $-$ & $\leq$~0.03 & $\leq$~2.5 & $-$ & 21.16~$\pm$~0.02 &  3.6~$\pm$~0.3 & 437/324\\
& & & & & & & \\          
\hline \noalign{\smallskip}
NQ Gem & \it{(partcov~$\times$~tbabs)(apec$_{1}$ + apec$_{2}$ + gaussian)} & 0.027~$\pm$~0.001 & cf=0.990~$\pm$~0.002 &  1:0.2~$\pm$~0.1 & 1.0 & 1:2.3$^{+0.9}_{-0.8}$ & 1:2.9$^{+1.2}_{-1.1}$ & 3427/1690\\ 
\\ 
& & $-$ & full=50$^{+6}_{-5}$ & 2:7.0$^{+1.6}_{-1.2}$ & 1.0 & 2:15.3$^{+0.9}_{-0.8}$ & 2:19$^{+4}_{-3}$ & $-$ \\
& & & & & & & \\  
\hline \noalign{\smallskip}
CD -27 8661 & tbabs~$\times$~(gaussian+apec)& 0.004~$\pm$~0.001 &  $\leq$~85 & 7$^{+10}_{-6}$ & 2.2~$\pm$~0.1  & 2.5~$\pm$~0.7  & 9$^{+3}_{-2}$ & 314/104\\
& & & & & & & \\  
\hline \noalign{\smallskip}
\end{tabular}\\
Notes:
$^{a}$ Indicates the value of the maximum temperature in the case of the cooling ﬂow model mkcﬂow, kT$_{max}$ and a plasma temperature for the model \textit{compTT}.\\ 
\end{sidewaystable*}

\begin{table*}
\caption{\label{bayes_analysis_priors} Priors minimum and maximum and the values found from BXA.}
\small
\centering
\begin{tabular}{cccccccc}
\hline\hline 
Object & Model & Name & Method & Min & Max  & Value & Units\\
\hline
\\
BD Cam & \it{tbabs~$\times$~apec} & N$_{H}$ &  log(LogUniform) & 10$^{-2}$  & 5 & -1.48$^{+0.49}_{-0.37}$ & 10$^{22}$~cm$^{-2}$\\
\\
       &            & kT     & Uniform  & 10$^{-5}$ & 20 & 2.80$^{+1.04}_{-0.70}$ & keV\\ 
\\
       &            & norm    &  log(LogUniform)  &  10$^{-8}$ & 1 &  -2.84$^{+0.06}_{-0.06}$ & $-$\\
\\
\hline\\
V1261 Ori & \it{tbabs~$\times$~compTT} & N$_{H}$ &  log(LogUniform) & 10$^{-2}$  & 5 & -1.70$^{+0.22}_{-0.20}$ & 10$^{22}$~cm$^{-2}$\\
\\
& & T0 & Uniform & 10$^{-2}$ & 10 & 0.18~$\pm$~0.01 & keV\\ 
\\
& & kT & Uniform & 10$^{-2}$ & 10 & 2.15~$\pm$~0.03 & keV\\ 
\\
& & norm & log(LogUniform) &  10$^{-8}$ & 1 & -3.25~$^{+0.03}_{-0.02}$ & $-$\\
\\
\hline\\
NQ Gem & \it{(partcov~$\times$~tbabs)(apec$_{1}$ + apec$_{2}$ + gaussian)} & CvrFract &  Uniform & 10$^{-8}$ & 1 & 0.99 & $-$\\
\\
& &  N$_{H}$ & log(LogUniform) & 10$^{-2}$ &  100 &  1.70~$\pm$~0.03  & 10$^{22}$~cm$^{-2}$\\
\\
& &  kT$_{1}$ & Uniform & 10$^{-2}$ &  10 & 0.13$^{+0.03}_{-0.02}$ & keV \\
\\
\\
& &  kT$_{2}$ & Uniform & 10$^{-2}$ &  10 & 7.16$^{+0.93}_{-0.82}$ & keV\\
\\
\\
& &  norm$_{Gaussian}$ & log(LogUniform) & $-$ & $-$ & -5.41$^{+0.09}_{-0.11}$ & $-$\\
\\
\hline\\
CD -27 8661 &  tbabs~$\times$~(gaussian+apec) & N$_{H}$ & log(LogUniform) &  10$^{-2}$  & 5.0 & 1.94$^{+0.18}_{-0.19}$ & 10$^{22}$~cm$^{-2}$\\
\\
& &  norm$_{Gaussian}$ & log(LogUniform) & $-$ & $-$ & -5.84$^{+0.29}_{-0.54}$ & $-$\\
\\
& &  kT     & Uniform   & 10$^{-5}$ & 20 & 11.43$^{+9.02}_{-4.10}$ & keV\\
\\ 
& &  norm & log(LogUniform) & $-$ & $-$ & -3.78$^{+0.26}_{-0.22}$ & $-$\\
\hline \\
\end{tabular}
\end{table*}

\begin{figure*}
\centering
\includegraphics[scale=0.4]{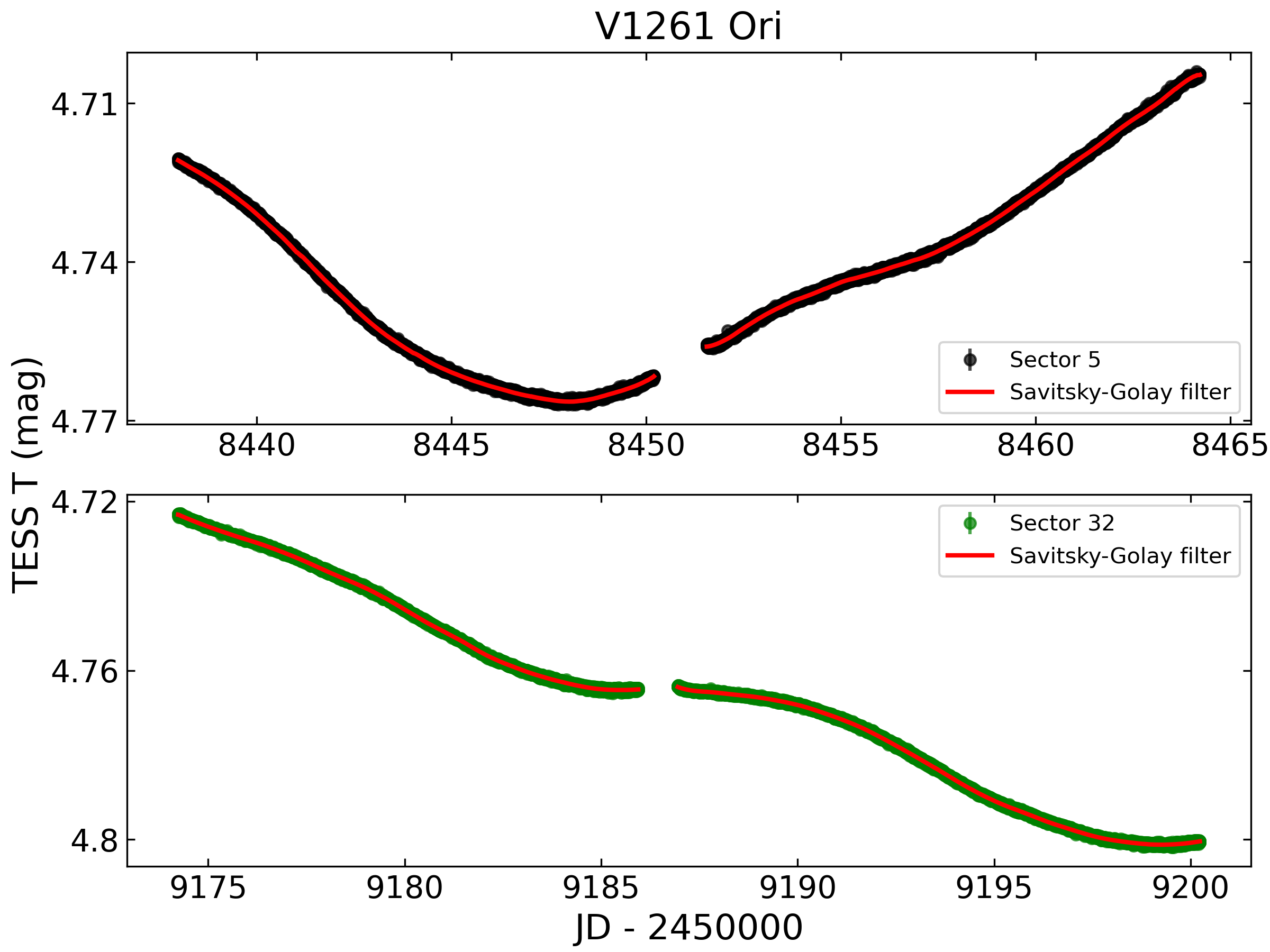}
\includegraphics[scale=0.4]{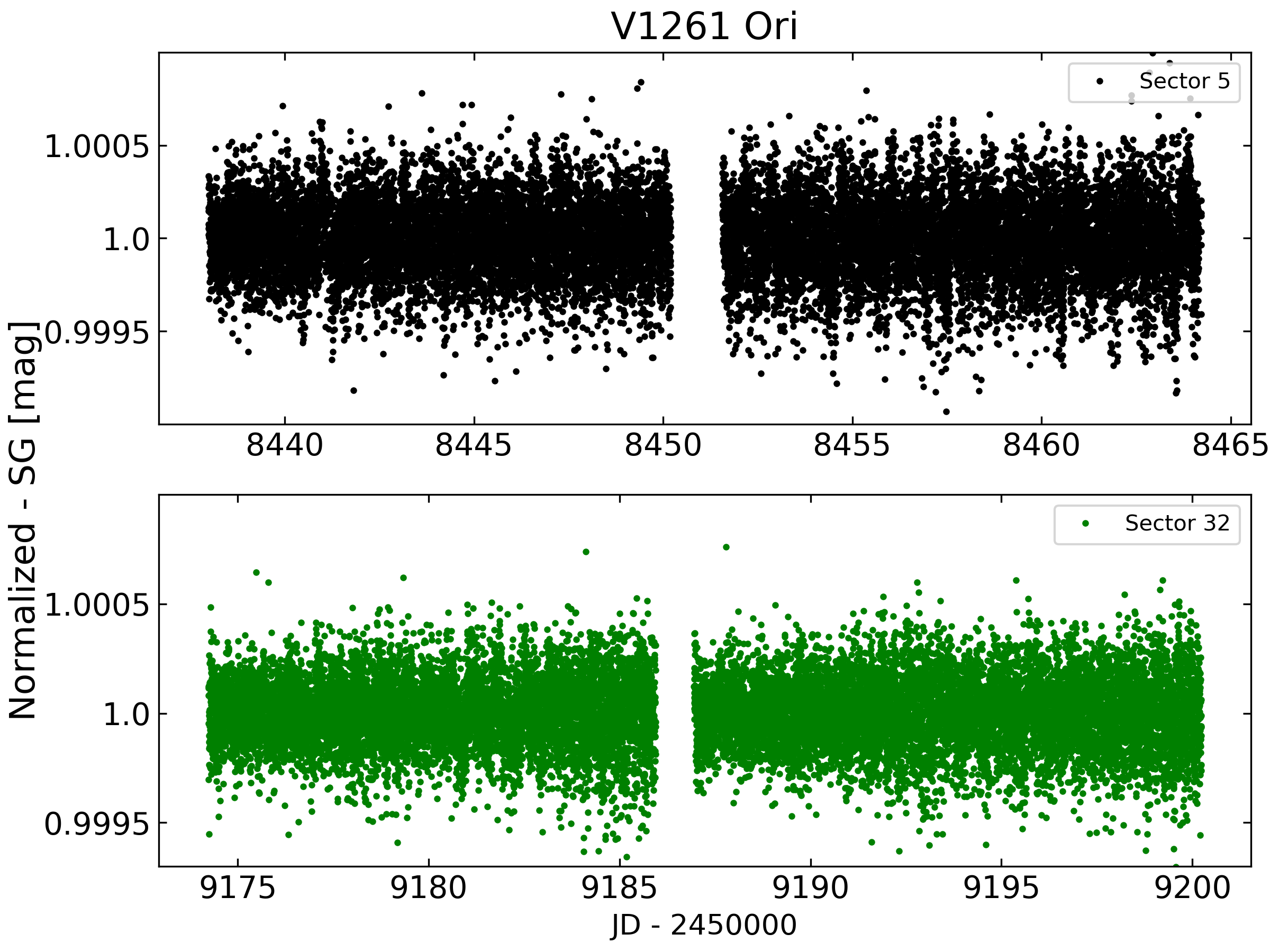}
\includegraphics[scale=0.4]{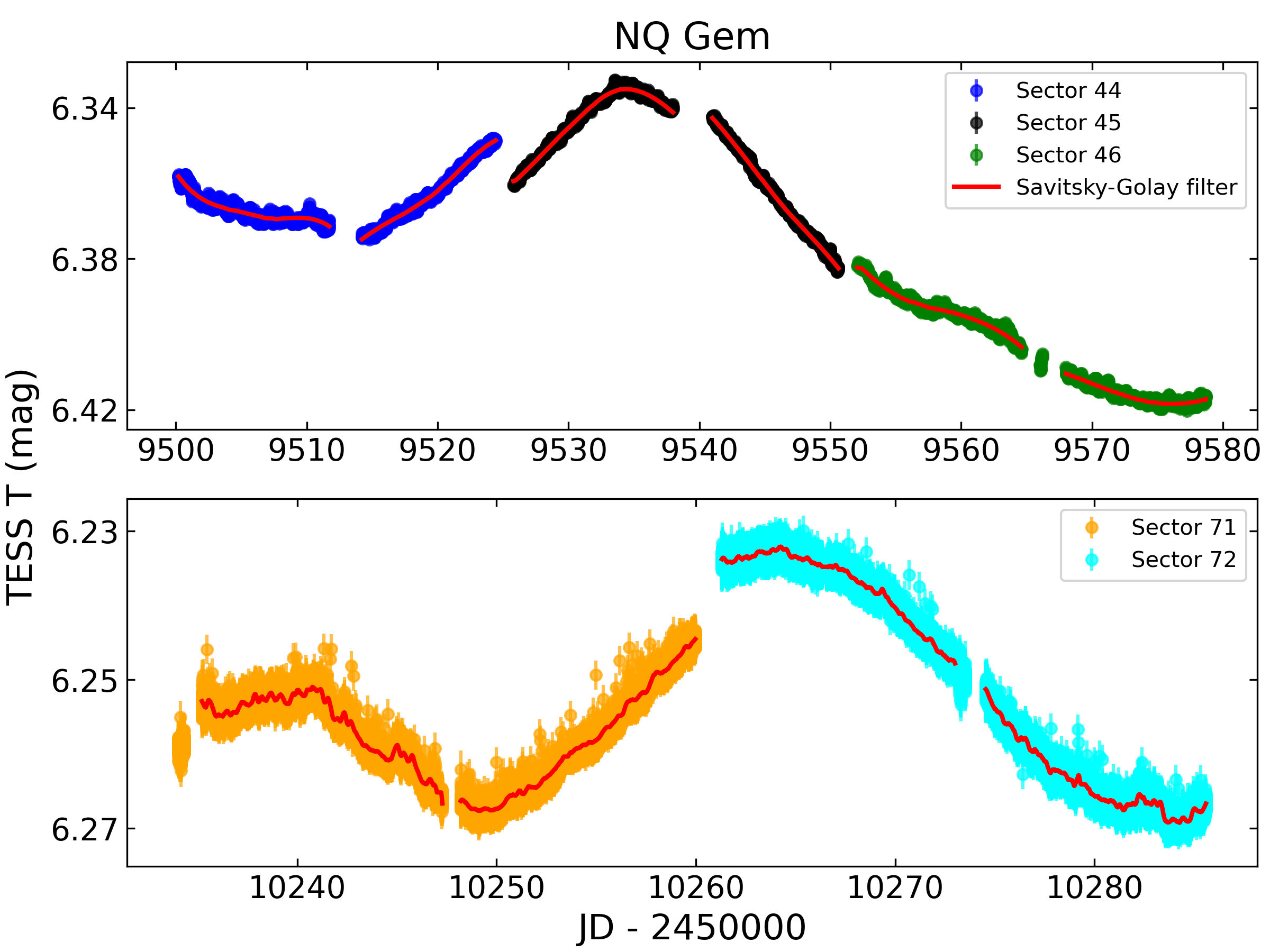}
\includegraphics[scale=0.4]{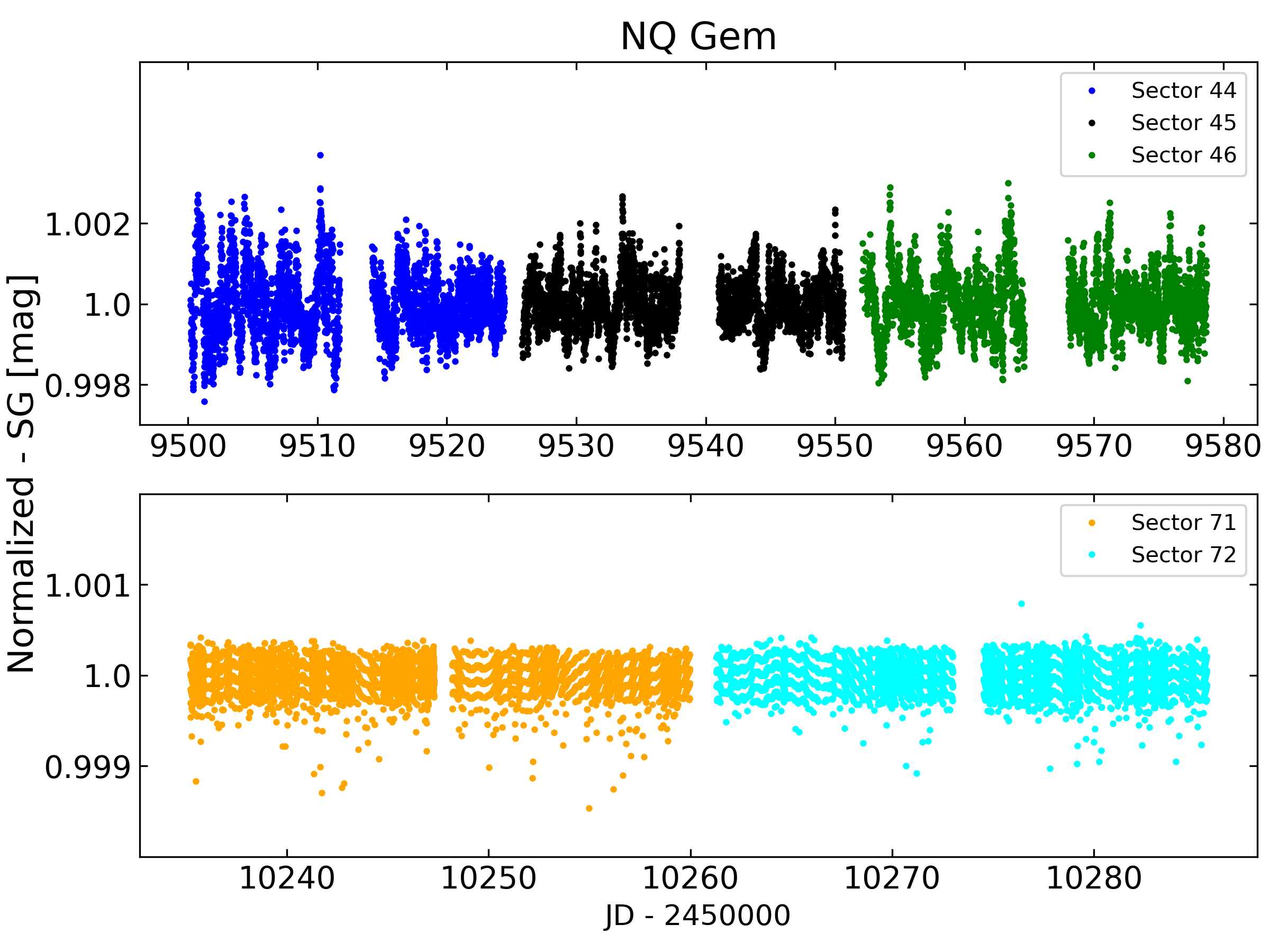}
\includegraphics[scale=0.4]{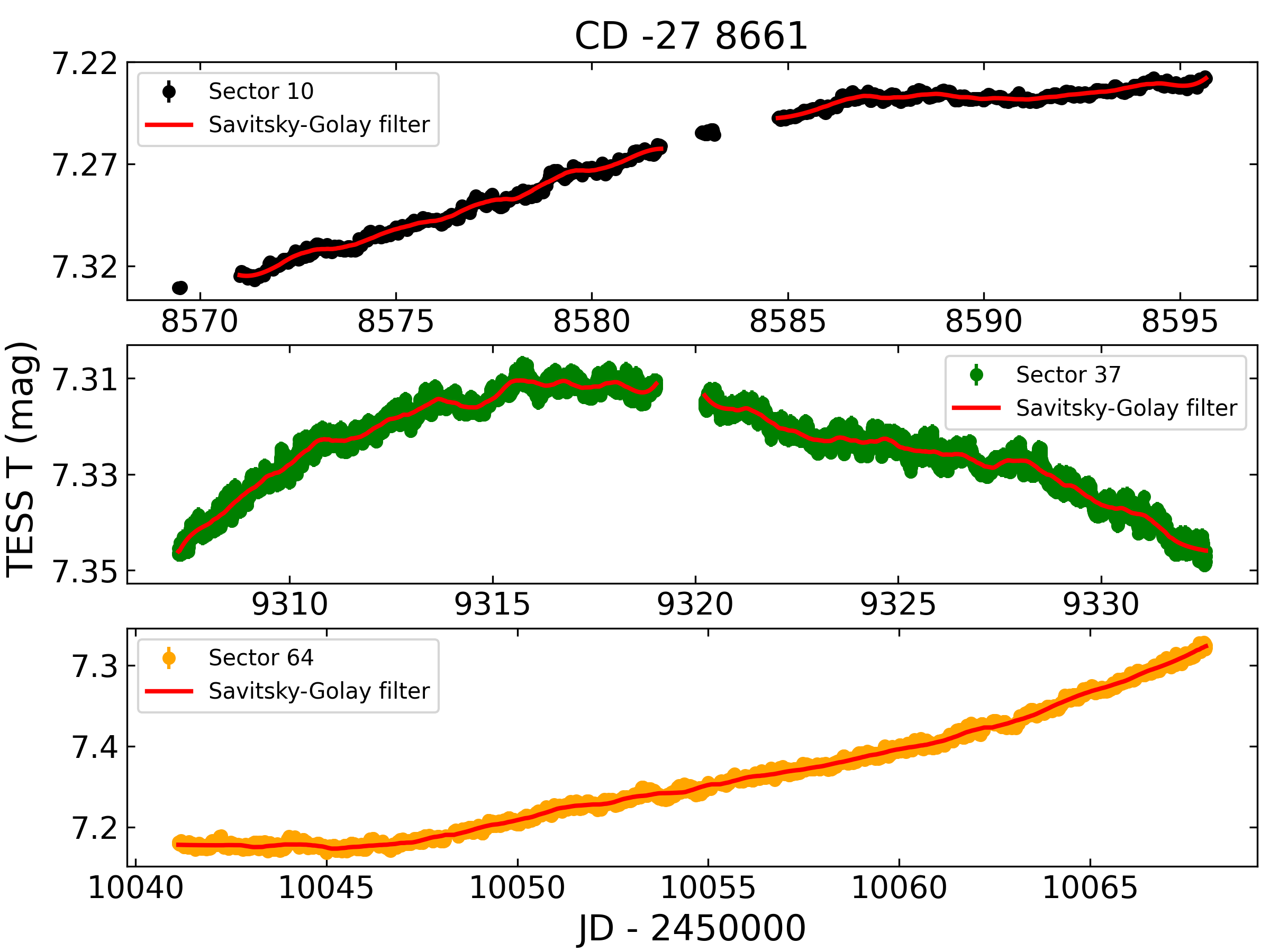}
\includegraphics[scale=0.4]{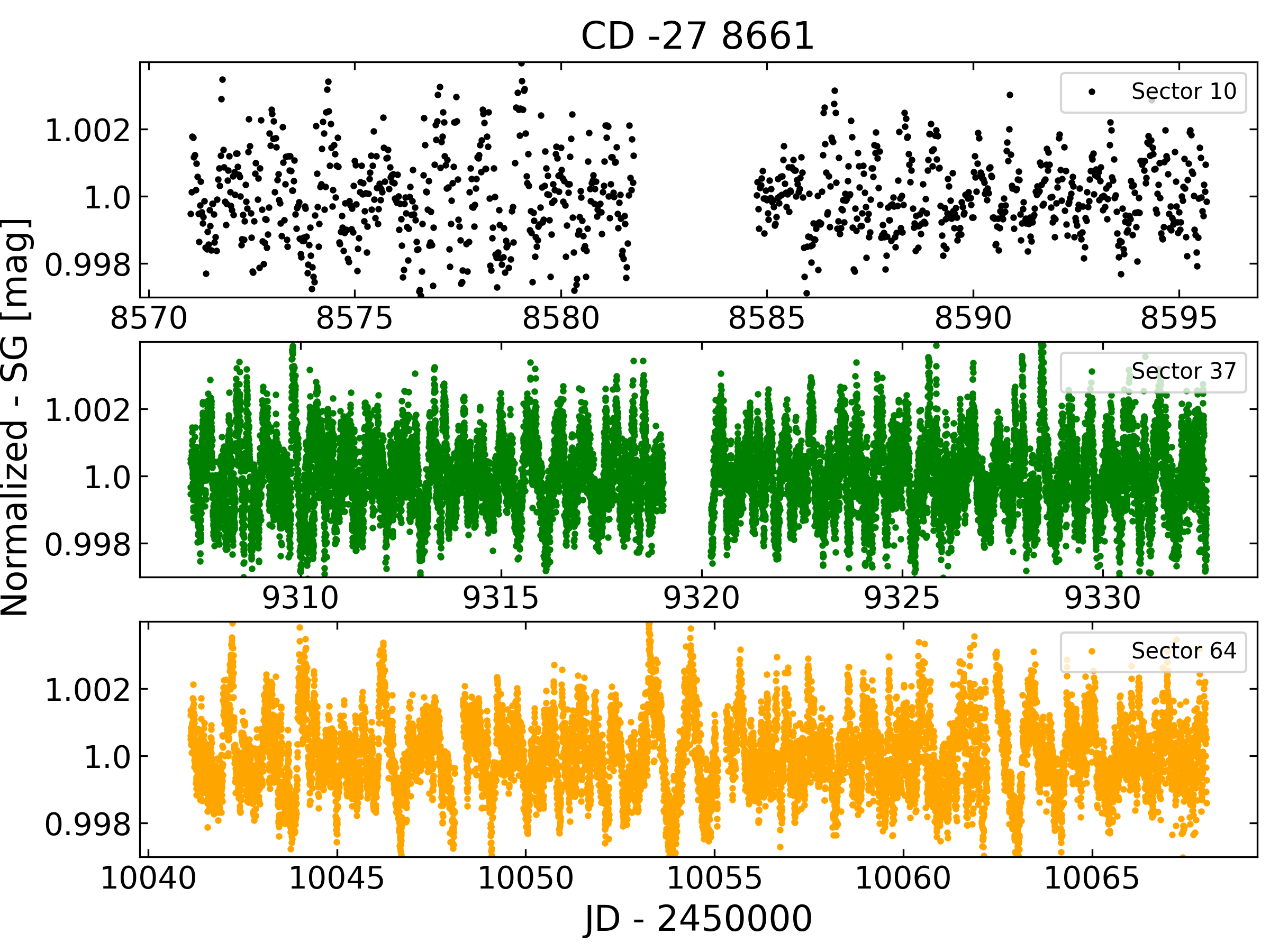}
\caption{Left column: \tess~light curves of V1261~Ori, NQ~Gem, and CD~-27~8661 for each sector with the Savitsky-Golay over-plotted in red. Right column: amplitude of normalized \tess\ T magnitude after subtracting the SG filter. The gaps are the data transmission to Earth.}
\label{fig:tess}
\end{figure*}

\subsection{BD Cam} \label{subsec:bd} 

BD~Cam is a symbiotic binary with a Tc-deficient and a red giant component with T$_{eff}$~=~3500~K (e.g., \citealt{Ake_1988}). 
It was discovered by \cite{Keenan_1954} as an S5.3 spectral type star with a magnitude of V~=~5.1~mag and $\sim$1~mag amplitude variations between high and low states (see Figure~\ref{fig:historical-light-curves}).
In their analysis of the IUE spectra, \citet{Ake_1988} showed that strong UV emission lines and the observed flux variability were related to a WD companion, thus establishing the symbiotic nature of the system.  
The system has an orbital period of 596~d \citep{Griffin_1984} and pulsation of 24.76~d associated to giant \citep{Adelman_1998}. 
The mass loss rate from the cool component, derived from dust-loss rate measurements from Two Micron Sky Survey data \citep{Jura_1988}, was estimated to be 1.7~$\times$~10$^{-8}$~M$_{\odot}$~yr$^{-1}$.

\cite{Ortiz_2019} found emission lines of OIV]/Si~IV$\lambda$1400, C~IV$\lambda\lambda$1548,1551 and OIII]$\lambda$1666 in GALEX and IUE far-UV spectra. Longer wavelength IUE
spectra revealed absorption lines from the red giant at $\lambda$~>~2800~\AA.
In an optical spectrum obtained with the Isaac Newton Telescope (INT), they found an emission core in the Ca~II absorption line, commonly associated with chromospheric activity. Occasionally the He~I~$\lambda$10830 line appears in absorption, suggesting that this line is formed in a layer above the photosphere of the primary, which is not expected under the hypothesis of an accretion disk around the secondary. 

The X-ray observations of BD~Cam with \swift\ were described in Section~\ref{subsec:swift}.
The X-ray spectrum was modeled in the range 0.6~--~9~keV due to high background counts contribution at lower energies. We fitted the spectrum with a simple model composed of an absorbed optically thin thermal plasma 
(\textit{tbabs~$\times$~apec}), with kT~=~2.3$^{+1.8}_{-0.5}$~keV (see Figure~\ref{fig:spectra-analises}). 
The absorption column was N$_{H}$~$<$~0.5~$\times$~10$^{22}$~cm$^{-2}$ (see Table~\ref{fig:spectra-analises}). The unabsorbed flux was F$_{X}$~=~18$^{+7}_{-1}$~$\times$~10$^{-13}$~erg~s$^{-1}$~cm$^{-2}$ and the luminosity was L$_{X}$~=~1.2$^{+0.5}_{-0.2}$~$\times$~10$^{31}$~erg~s$^{-1}$ at a distance of 235~$\pm$~15~pc given in the \gaia\ Early Data Release 3 (EDR3, \citealt{Bailer-Jones_2021}). 

BD~Cam was previously classified as $\beta$/$\delta$-type by \cite{Merc_2021} based on the X-ray spectrum from the \swift-XRT point source catalog by \cite{Evans_2013}. However, as mentioned in Section~\ref{subsec:swift}, the XRT data are heavily affected by optical loading, with the appearance of artificial soft and hard components in the spectrum (see Figure~\ref{optical_loading}). The free-from-optical-loading spectrum taken in WT mode shows that the majority of the photons have energy lower than 2.4~keV, while the plasma temperature reaches values of 2.3 keV (see Table \ref{tab:spectral-models}). If this temperature arises from shocks in a colliding wind region, the strong shock conditions imply that the shock speed is about 1500 km~s$^{-1}$ (e.g., \citealt{nunez_2016}). Given that such speeds are not reported for BD~Cam, we favor a scenario where those speeds are present in the shock within the accretion disk boundary layer, and thus BD~Cam should be classified as a $\beta$- or $\delta$-type symbiotic star. 

The Bayesian analysis of X-ray data from BD~Cam using the same model described above, yielded a temperature of $kT$~=~2.80$^{+1.04}_{-0.70}$~keV considering the absorption column density of 0.03~$\times$~10$^{22}$~cm$^{-2}$ (see Table~\ref{bayes_analysis_priors}). These results agree with the nonBayesian approach. 
Figure~\ref{bxa_results_corner} shows the covariances of the posterior distributions of the free parameters N$_{H}$, $kT$, and normalization and its values. These results indicating that the model globally reproduces the data. 

\subsection{V1261 Ori} \label{subsec:v12}

V1261~Ori is the first eclipsing binary star discovered with a cool barium component, of spectral type S4.1 (T$_{eff}$~=~3470~K). A strong Ce~II absorption line was identified in the spectra from the APOGEE survey indicating the presence of $s$-process elements due to AGB stellar evolution \citep{Cunha_2017}.
The symbiotic star classification of V1261~Ori is based on the presence of high-excitation emission lines (e.g., C~IV~$\lambda$~1550~\AA) and a variable continuum in the UV, both being typical of interacting binaries \citep{Ake_1991, Belczynski2000}.
A modulation of 640.5~d was reported by \cite{Gromadzki_2013} from V-band ASAS-SN data, in the form of a double sine-wave, and has been interpreted as due to ellipsoidal variations. Later, 
\citet{Boffin_2014} reported a period of 
638.24~$\pm$~0.28~d interpreted as due to the orbital modulation. 
\cite{Boffin_2014} has argued against the presence of ellipsoidal variations based on the red giant's small filling-factor of 0.32~$\pm$~0.16, and on the absence of synchronization between the spin of the red giant and the orbital period. 

V1261~Ori presents characteristics from red and yellow-type symbiotic stars: the low metal abundance of post–iron-peak elements is typical of yellow symbiotics, while its luminosity and temperature are commonly observed in red symbiotics. However, the mass-loss rate of 10$^{-8}$ M$_\odot$~yr$^{-1}$ determined by \cite{Jorissen_1996} does not agree with the mass-loss rates of either red or yellow systems, typically around 10$^{-7}$~M$_\odot$~yr$^{-1}$. There is no clear explanation for this reduced mass-loss rate, but it should lead to a lower density of circumstellar material around the system and may account for the absence of optical emission lines \citep{Vanture_2003}.

The \swift\ data of V1261~Ori suffered from optical loading as described in Section~\ref{subsec:swift}, hence we analyzed only the \xmm~data, which were obtained using the thick filter in the EPIC camera and thus avoided contamination by optical loading. 
We considered two spectral models for V1261~Ori, an absorbed multitemperature cooling-flow plasma (\texttt{tbabs~$\times$~mkcflow}) and an absorbed comptonized spectrum (\texttt{tbabs~$\times$~compTT}). The choice of this latter model, usually used to describe the X-ray emission of $\gamma$-type symbiotics harboring a neutron star, was motivated by the apparent absence of emission lines, which should be present in an optically thin thermal plasma.
For the first model, we found an absorbing column of N$_{H}$~=~0.07~$\pm$~0.01~$\times$~10$^{22}$~cm$^{-2}$, a low temperature of $kT$~=~0.1~$\pm$~0.05~keV, and a high temperature of $kT$~=~ 7.7~$\pm$~0.3~keV (see Table~\ref{tab:spectral-models}). The accretion rate derived from the model normalization was 2.82$^{+0.18}_{-0.15}$~$\times$~10$^{-11}$~M$_{\odot}$~yr$^{-1}$ \citep{Mushotzky_1988}.
The spectrum does not exhibit significant emission lines and the abundance was 0.17~$\pm$~0.04 using the abundance table of \texttt{wilm} \citep{Wilms_2000}. In fact, the [Fe/H] abundance derived from optical data is low, a value of -0.8~$\pm$~0.04 \citep{Vanture_2003}, that would be a supporting evidence for this low abundance obtained by \texttt{mkcflow} fit.
In the case of the second model, we found an absorber with a column density of N$_{H}~\leq$~0.03~$\times$~10$^{22}$~cm$^{-2}$, 
a soft photon (Wien) temperature of 0.18~$\pm$~0.01~keV, a plasma temperature of $\leq$~2.5~keV, and the optical depth of the plasma cloud of 6.4$^{+0.4}_{-0.6}$. 
The unabsorbed fluxes are quite similar for both models: 23.71~$\pm$~0.02 and 21.16~$\pm$~0.02~$\times$~10$^{-13}$~erg~s~$^{-1}$~cm$^{-2}$. The luminosities are 3.9~$\pm$~0.3 and 3.6~$\pm$~0.3~$\times$~10$^{31}$~erg~s~$^{-1}$ using \gaia's distance of 375~$\pm$~17~pc.
 
The Bayes X-ray spectral analysis was applied only to the pn data of V1261~Ori. We also fixed the abundance of \texttt{mkcflow} model and the plasma optical depth of \texttt{compTT} model. The posterior distributions for the \texttt{tbabs~$\times$~mkcflow} model showed uniformly flat values over all parameter space, its results mostly follow the priors (see Figure~\ref{bxa_results_corner} in Appendix~\ref{appendix2}). This model is not able to recover the sensible parameters of multitemperatures ($lowT$, $highT$) and absorption (N$_{H}$). On the other hand, the \texttt{tbabs~$\times$~compTT} model suggested that the plasma has temperatures of 0.18~$\pm$~0.01~keV (Wien) and 2.15~$\pm$~0.03~keV with absorption $\leq$~0.02~$\times$~10$^{22}$~cm$^{-2}$.
Table~\ref{bayes_analysis_priors} shows these values. 
These results are similar to the values found using nonBayes analysis. 

\begin{figure}
\centering
\includegraphics[scale=0.28]{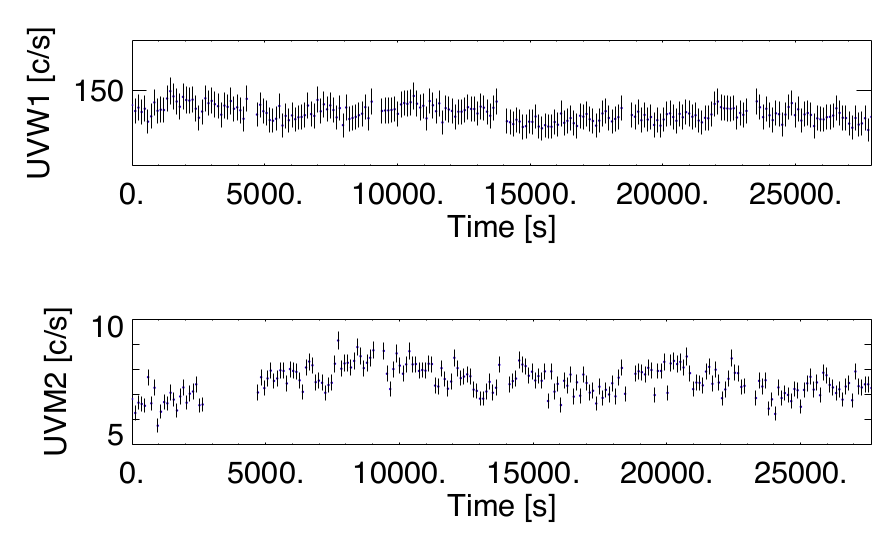}
\caption{The \xmm-OM light curves of V1216~Ori in the UVW1 and UVM2 filters in 120~s bins.}
\label{fig:OM_v1261}
\end{figure}

V1261~Ori was observed with \tess\ in two sectors, 5 and 32, separated by about two years. The earliest observation was performed when the source was at an average brightness level of about 6.8~mag, as shown in AAVSO V band data (see Figure~\ref{fig:historical-light-curves}). 
From \tess\ data, neither sectors 5 and 32 show a presence of modulations (see Figure~\ref{fig:tess}). We found 
flickering amplitudes of  0.02\% for each four parts of the \tess\ light curves (see Table~\ref{tab:flickering}).
At shorter wavelengths, the UVW1 and UVW2 data from \xmm-OM exhibit variability of about one tenth of a magnitude (see Figure~\ref{fig:OM_v1261}) with average amplitude flickering of 3.6\%. 

\subsection{NQ Gem} \label{subsec:nq}

The red giant component in NQ~Gem is a carbon star of spectral type C6.2 \citep{Yamashita_1972}. It was first classified as a symbiotic star by \cite{Greene_1971} due to its highly variable UV continuum and the similarity of its optical spectra with the symbiotic recurrent nova T~CrB. Observations with the IUE satellite showed strong C~IV~$\lambda$1550 emission line, and Si~III]/C~III]~$\lambda$1900 lines ratio similar to those of symbiotic stars 
\citep{Johnson_1988}. The orbital period of 1305~$\pm$~4~d was found by \cite{Carquillat_2008} and a pulsation period of 58~$\pm$~1~d was reported by \cite{Gromadzki_2013}.

NQ~Gem has been classified as a $\beta$/$\delta$ X-ray type symbiotic. \cite{Luna_2013} reported the X-ray spectrum obtained with \swift\ that exhibited two distinct components at energies above and below $\sim$1.5~keV; a similar X-ray spectrum has been observed in a few WD symbiotics (e.g., CH~Cyg, ZZ~CMi, V347~Nor, and UV~Aur). The X-rays with energies above 2~keV presumably arise from the optically thin portion of the accretion disk boundary layer, while those X-rays with lower energies could be generated in a region where the winds from the red giant and that from the white dwarf/accretion disk
\cite{Luna_2013} modeled both optically thin thermal components with temperatures of kT~=~0.23~$\pm$~0.03~keV and kT~$\gtrsim$~16~keV for soft and hard energies, respectively. The unabsorbed flux in 0.3~-~10~keV was 6.7$^{+0.9}_{-0.8}$~$\times$~10$^{12}$~erg~s~$^{-1}$~cm$^{-2}$ and a luminosity of 80$^{+1.1}_{-0.9}$~$\times$~10$^{31}$~erg~s~$^{-1}$ was estimated, at distance of 1~kpc.

The \xmm\ data were obtained during a typical brightness state with magnitude of about 8~mag in V band (see Figure~\ref{fig:historical-light-curves}). We applied a spectral model similar to the one used by \cite{Luna_2013} which is composed by one partial covering absorber, two thermal components, and a Gaussian line profile (see Table~\ref{tab:spectral-models}). It yielded a good fit to the soft and hard X-ray emission. Our analysis of the \xmm~spectra yielded N$_{H}$(partial)=50$^{+6}_{-5}$~$\times$~10$^{22}$~cm$^{-2}$ with partially covering fraction of 0.990~$\pm$~0.002, and thermal components of kT$_{1}$~=~0.2~$\pm$~0.1~keV and kT$_{2}$~=~7.0$^{+1.6}_{-1.2}$~keV.  

\cite{Toala_2023} modeled the same \xmm~data with two absorbed temperature components at energies below 4~keV, a power law, and a heavily-absorbed thermal component with a Gaussian line profile, at energies above 4.0 keV. However, we did not find a significant improvement by including a second \textit{apec} component neither a power law to fit the 2.0~-~4.0~keV energy range. Moreover, the presence of nonthermal emission in the X-ray spectra of symbiotics is yet to be conclusively established and thus the physical motivation to include it in the spectral models is still marginal. 

We found unabsorbed fluxes of 2.3$^{+0.9}_{-0.8}$~$\times$~10$^{-13}$~erg~s~$^{-1}$~cm$^{-2}$ and 15.3$^{+0.9}_{-0.8}$~$\times$~10$^{-13}$~erg~s~$^{-1}$~cm$^{-2}$ for the soft and hard components, respectively. The most up-to-date distance of NQ~Gem is 1012$^{+99}_{-76}$~pc from \gaia\ (EDR3), which implies luminosities of 2.9$^{+1.2}_{-1.1}$~$\times$~10$^{31}$~erg~s~$^{-1}$ and 19$^{+4}_{-3}$~$\times$~10$^{31}$~erg~s~$^{-1}$. Our results are in agreement with those presented in \cite{Luna_2013} using \swift\ data and \cite{Toala_2023} using \xmm\ data.

The Bayesian X-ray analysis was applied only for pn data, since the EPIC-MOS data presented low number of counts. We also fixed the normalization of the two thermal models with values found from C-statistic. Therefore, the free parameters were the covering fraction, absorption, temperatures for soft and hard components, and normalization of the Gaussian. We found N$_{H}$~=~50.1~$\pm$~1.1~$\times$~10$^{22}$~cm$^{-2}$ with a covering fraction of 0.99, $kT$~=~0.13$^{+0.03}_{-0.02}$~keV and 7.16$^{+0.93}_{-0.82}$~keV for soft and hard components, respectively (see Table~\ref{bayes_analysis_priors}).   
These values are consistent with our previous model using C-statistic approach.

NQ~Gem was observed with \tess\ in three consecutive sectors, covering a period of almost three months in 2021, and recently in October and November 2023. {\it AAVSO}-V band photometry during those months in 2021 shows that NQ~Gem was about 8.3 mag (see historical data in Figure~\ref{fig:historical-light-curves}). 

\begin{figure}
\centering
\includegraphics[scale=0.33]{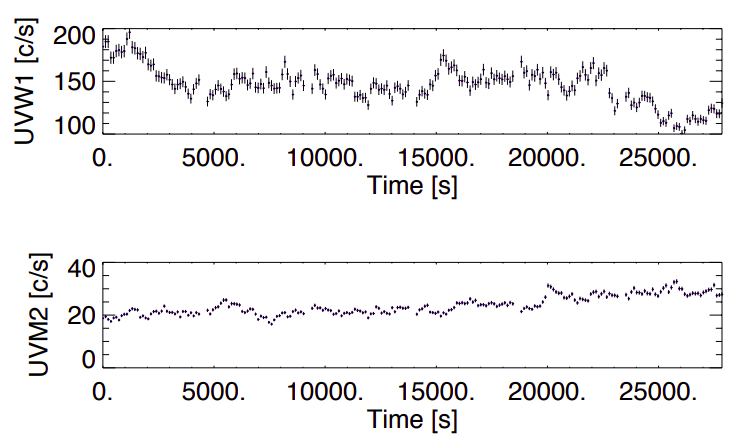}
\caption{The \xmm\ OM light curves of NQ~Gem in the UVW1 and UVM2 filters in 120 s bins.}
\label{fig:OM_nqgem}
\end{figure}

\cite{Luna_2013} reported flickering 
with fractional rms amplitude of 15\% from \swift~UVOT photometry. Indeed, we found about 7\% of flickering amplitude for \xmm-OM light curves in the UVW1 and UVM2 filters (see Figure~\ref{fig:OM_nqgem}).
During \tess\ observations, the average amplitude of variability was 0.1\% for sectors 44, 45, and 46, while during sectors 71 and 71 was 0.01\%, these later light curves are more noise than its early data. Figure~\ref{fig:tess} shows the \tess\ time series of sectors 44, 45, 46, 71 and 72 with typical gaps used for data transfer to Earth (left column) and the amplitude of flickering after remove the SG fit (right column).  

\subsection{CD -27~8661} \label{subsec:cd27}

CD -27~8661 (V420~Hya) is another eclipsing, extrinsic S symbiotic star with no presence of Tc element in the red giant spectrum \citep{Van_Eck_2002}. 
Its symbiotic nature was suggested based on a strong blue-violet continuum and broad H$\alpha$ emission with FWHM~$>$~300~km~s$^{-1}$ \citep{Van_Eck_2000}. The authors also pointed out that H$\alpha$ originates from the gas moving with the companion star, because the radial velocity from the center of the emission line seems to be similar to that of the companion star. \cite{Van_Eck2000a} found that the spectral features such as the flux of the UV continuum and the H$\alpha$ emission line are related with the orbital period of 763.63~$\pm$~5.88~days. This orbital period is roughly in agreement with the value of 751.4~$\pm$~0.2~d previously proposed by \cite{Jorissen_2019}.  
The secondary is close to fill its Roche lobe as the calculated secondary radius is 153~R$_{\odot}$ while its Roche lobe radius is 180~R$_{\odot}$ \citep{Van_Eck_2002}. The authors also estimated an
orbital eccentricity of 0.099~$\pm$~0.004. 

CD -27~8661 was observed on 2010 August 24 and 26
by \swift. The \swift~UVOT light curves were presented by \cite{Luna_2013} with a mean count rate of 17.7~$\pm$~0.5~c~s$^{-1}$. 
We analyze here the \xmm~observations performed on 2016 January 27, during orbital phase 0.92 (using the ephemeris from \citealt{Jorissen_2019}) and therefore out of the eclipse. The X-ray spectrum was modeled in the 4 to 10~keV energy range using an absorbed optically thin thermal plasma plus a Gaussian emission line which gives a statistically acceptable description of the observed spectrum. We found an absorbing column of N$_H$~$\leq$~85~$\times$~10$^{-22}$~cm$^{-2}$, a temperature of 7$^{+10}_{-6}$~keV, 
and metal abundance of 2.2~$\pm$~0.1. The center of the Gaussian emission line was fixed at 6.4~$\pm$~0.005~keV. The unabsorbed flux was 2.5~$\pm$~0.7~$\times$~10$^{-13}$~erg~s~$^{-1}$~cm$^{-2}$, corresponding to a luminosity of 9$^{+3}_{-2}$~$\times$~10$^{31}$~erg~s~$^{-1}$ at a distance of 1679$^{+148}_{-119}$~pc obtained from \gaia\ (EDR3). Based on its hard X-ray emission, with most of the emission with energies above 4~keV, CD~-27~8661 can be classified as $\delta$-type symbiotic. 

We also obtained the previous model parameters using the Bayesian methodology. In the energy range of 4 to 10~keV, we found the posteriors N$_H$~=87.1$^{+1.5}_{-1.6}$~$\times$~10$^{-22}$~cm$^{-2}$ and kT~=~11.4$^{+9.0}_{-4.1}$~keV  (see Table~\ref{bayes_analysis_priors}). 
The abundance was fixed at 2.2 and the Gaussian emission line center and width were also fixed as the model was insensitive to their precise values. In turn, the normalization of the \textit{Gaussian} and \textit{APEC} models were considered free parameters. These resulting values for the free parameters are in agreement with those found using only C-statistic.

\begin{figure}
\centering
\includegraphics[scale=0.33]{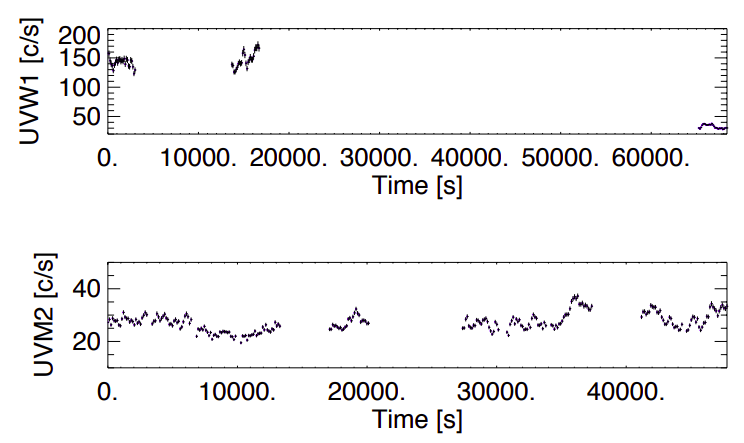}
\caption{The \xmm\ OM light curves of CD~-27~8661 in the UVW1 and UVM2 filters with 120 s bins.}
\label{fig:OM_cd27}
\end{figure}

The \tess\ observations of CD~-27~8661 were carried out in three sectors, 10, 37, and 64 separated by two years between them. The light curves are shown in Figure~\ref{fig:tess} (left panel). Subtracting the light curves from SG fitting, see right panel in Figure~\ref{fig:tess}, flickering activity was detected with amplitude of 0.2\% in sector 10, 0.1\% in sectors 37 and 64. Figure~\ref{fig:OM_cd27} shows the \xmm-OM light curves from UVW1 and UVM2 filters with flickering amplitude slightly higher around 8\%.  

\section{Discussion} \label{discussion}

The symbiotic stars were historically known as soft X-ray emitters. The hard X-ray observatories such as \integral\ and \swift\ revealed that these systems can also be hard X-rays sources. The detection of hard X-rays from symbiotics not only provided a growing numbers of members of the class but also required a 
new type within the classification scheme proposed by \cite{Muerset_1997}. 
In this paper we find that, in terms of X-ray classification, 
BD~Cam can be a $\beta$- or $\delta$- type system, V1261~Ori and CD~-27~78661 can be $\delta$-type, and we confirmed the previous classification of NQ~Gem as $\beta$/$\delta$-type \citep{Luna_2013, Toala_2023}. 
We remark that 
V1261~Ori was already classified as $\delta$-type by \cite{Merc_2021} during the analysis of the 
\swift/XRT-PC mode data, which were affected by optical loading as described in Section \ref{subsec:swift}. 

\subsection{X-ray spectral analysis} \label{xanalysis}

Our best-fit models to the X-ray spectra consist of optically thin thermal emission seeing through an intrinsic absorber. 
Initially, we found the best-fit model parameters by using a C-statistic maximum-likelihood function. 
We then applied the Bayesian approach from the BXA package \citep{Buchner_2014}. The main goal was to distort the parameters space by priors in order to identify the posterior distribution of each parameter in multidimensional space, compiling the results in a Gaussian approximation. 
Table~\ref{bayes_analysis_priors} shows the intervals and the values of the parameters found by Bayes statistic. 
The results are quite similar to those obtained using C-statistic. For example, for NQ~Gem we found the absorber column density of N$_{H}$~=~50$^{+6}_{-5}$~$\times$~10$^{22}$~cm$^{-2}$ using nonBayes and N$_{H}$~=~51$~\pm~1$~$\times$~10$^{22}$~cm$^{-2}$ using Bayes method, that are congruent within the uncertainties. Similarly, the Bayesian analysis of the symbiotic system RT~Cru performed by \cite{Danehkar_2021} showed the same temperature and absorbing column density found by \cite{Luna_2007}, in which the Bayesian approach was not applied.

Even if we did not use as priors the values for the parameters obtained from the C-statistic approach, but a range of values, we found that the agreement with the results yielded by the Bayes approach is independent of our procedure. As a test, we provided a range of priors, instead of a fixed value, and the results found with Bayes statistic were consistent within the uncertainties to those determined by {\texttt xspec} error command. The Kullback-Leibler ($KL$) divergence characterizes how strongly the posterior is influenced by the prior \citep{Kullback_1951}. The difference between the prior and posterior measure revealed a mean value of $KL$~=~0.46~$\pm$~0.06~bans, which shows than the posteriors are not dominated by the priors (see Figure~\ref{bxa_results_corner}). 

In V1261~Ori, the {\texttt mkcflow} model did not converge due to a higher degeneracy existent between the multitemperatures and the other parameters, in this case {\texttt comppTT} model better describe its X-ray spectrum. 
In general cases, the N$_{H}$ posterior does not introduce a strong bias, as mentioned by \cite{Buchner_2014}, although the majority of sub-volumes probability are slighter blurred, the classification of the sources among the absorbed and unabsorbed categories can be considered accurate.

In general, there is a degeneracy between N$_{H}$ and temperature parameters showed by the flat distribution of the parameters space (see the \textit{corner} module on the left panel in Figure~\ref{bxa_results_corner}).
Hence, a prior value of $N{_H}$ plays an important role in constraining $kT$. 
As mentioned before, a good fit would produce a straight line in Q~-~Q plots (right panel in Figure~\ref{bxa_results_corner}). In V1261~Ori using for the \texttt{compTT} model, we found $K~-~S$~value of 0.006. This value is close to 0 indicating a good fit.
In the case of BD~Cam and NQ~Gem, the red line is close to the gray line, indicating an statistically acceptable model with $K~-~S$ values of 0.097 and 0.040, respectively. The X-ray spectrum of CD~-27~8661 has a low numbers of counts, consequently showing the red line in saw-shaped a little distant from the gray line and $K~-~S$~=~0.071. This same behavior is also observed using C-statistic.

\subsection{Mass accretion rate of optically thin boundary layer} \label{rate}

The mass accretion rate ($\dot{M}_{thinBL}$) can be estimated by measuring the unabsorbed hard X-ray luminosity ($L_{x}$) between 0.3 and 10~keV\, which is related to the optically thin portion of the boundary layer,
and following its relationship with the white dwarf mass as (e.g., \citealt{Luna_2007}):

\begin{equation}
    \dot{M}_{thinBL} = \frac{2 L_{x}R_{WD}}{GM_{WD}}.
\label{eq_mass_accretion}
\end{equation}
\noindent where $L_{x}$ is the unabsorbed X-ray luminosity. $R_{WD}$ and $M_{WD}$ are the radius and the mass of the white dwarf, $G$ is the gravitational constant. 

Symbiotic systems have WD mass around 0.8~M$_{\odot}$ \citep{2008ASPC..401...42M}, although some symbiotic recurrent novae have WD masses close to the Chandrasekhar limit such as T~CrB, with 1.2~-~1.37~M$_{\odot}$ (e.g., \citealt{Luna_2018}). 

Assuming optically thin X-ray emission, and using the typical value of WD mass and WD radius of 7~$\times$~10$^{8}$~cm as well as the luminosities found from each spectral model (using the \gaia-DR3 distances), the mass accretion rate of our targets are given in Table~\ref{tab:mass_accretion_rate}.
In NQ~Gem, the mass accretion rates were calculated using the luminosities obtained for hard components, which is related to the accretion disk boundary layer.

As mentioned before, the normalization of cooling flow model yielded the mass accretion rate. In V1261~Ori, we found the value of $\sim$3~$\times$~10$^{-11}$~M$_\odot$~yr$^{-1}$ (see Section~\ref{subsec:v12}). This last result is consistent with 0.8~$\times$~10$^{-11}$~M$_{\odot}$~yr$^{-1}$ obtained using average values for the WD mass and radius (Equation~\ref{eq_mass_accretion}). At the same time, those values are significantly smaller than the values for the mass loss rate of Barium-syndrome stars such as BD~Cam and V1261~Ori proposed by \citealt[in the range of 10$^{-7}$~to~10$^{-8}$~M$_{\odot}$~yr$^{-1}$]{Jorissen_1996}. If the X-ray emission arises from the innermost region of the accretion disk, the low mass accretion rate inferred implies that this region is optically thin to its own radiation \citep{Patterson_1985}. On the other hand, the ratios of unabsorbed X-ray luminosity and the optical luminosity at filters UVW1 and UVM2 obtained from the OM data, which we assumed to have come from the Keplerian portion of the accretion disk, are extremely low (see Table~\ref{tab:luminosity_ratio}) indicating that the boundary layer is almost entirely optically thick in the X-rays with an optically thin layer that radiates the thermal bremsstrahlung component with temperatures smaller than about 8~keV \citep{Patterson_1985,Pandel_2005}. This suggests that the mass transfer mechanism in these systems is extremely inefficient when compared to the expectations from Roche-lobe filling or Bondi-Hoyle accretions. 

The historical optical light curves of our targets do not show significant brightness changes that could indicate a sudden increase in the mass accretion rate (see Figure~\ref{fig:historical-light-curves}) as those in novae or dwarf novae outbursts, which are more frequent in sources with high $\dot{M}$. This is in agreement with the low mass accretion rates that we have found in our analysis of the X-ray emission. 

\begin{table}
\caption[]{The mass accretion rate of thin optically boundary layer calculated to V1261~Ori, NQ~Gem, and CD~-27~8661.} 
\centering
\label{tab:mass_accretion_rate}
\begin{tabular}{cccccc}
\hline\hline  
\noalign{\smallskip}
Objects  &  Mass accretion rate  \\
& (10$^{-11}$~M$_{\odot}$~yr$^{-1}$)\\
\hline
\noalign{\smallskip}
V1261~Ori     &  0.8 \\
NQ~Gem        &  4.0 (hard) \\
CD~-27~8661   &  2.0 \\
\hline
\end{tabular}
\end{table}

\begin{table}
\caption[]{Luminosity relation of V1261~Ori, NQ~Gem, and CD~-27~8661.} 
\centering
\label{tab:luminosity_ratio}
\begin{tabular}{cccccc} 
\hline\hline  
\noalign{\smallskip}
Objects  & Filter &L$_{OM}$ & L$_{x}$/L$_{OM}$ \\
& & erg~s$^{-1}$ & \\
\hline
\noalign{\smallskip}
V1261~Ori & UVW1 & 8.3~$\times$~10$^{+32}$ & 0.05 \\
    &  UVM2 & 1.3~$\times$~10$^{+32}$ & 0.31\\
\hline  \noalign{\smallskip}         
NQ~Gem  &  UVW1 & 6.2~$\times$~10$^{+33}$  & 0.03 (hard)\\
& UVM2 & 3.0~$\times$~10$^{+33}$ & 0.06 (hard)\\
\hline \noalign{\smallskip}
CD~-27~8661 &  UVW1 &1.3~$\times$~10$^{+33}$ & 0.01\\
            & UVM2 & 9.1~$\times$~10$^{+33}$ & 0.01 \\
\hline
\end{tabular}
\end{table}
\subsection{Flickering variations} \label{flick}

Table~\ref{tab:flickering} lists the amplitude of the flickering for optical light curves from \tess~and \xmm-OM. 
We observe ﬂickering with small amplitude, ranging from 0.01 to 0.1\% of \tess\ data. \cite{Merc2023} found an average amplitude about 0.1\% for CD~-27~8661 and 0.03\% for NQ~Gem using the rms method (see their Figure~18). Our average amplitudes are similar for those targets, although NQ~Gem exhibits slight variation between sectors~44, 45, and 46 to 71 and 72.
The lower flickering amplitude observed in \tess\ data when compared to that from bluer bands such as the \xmm-OM and \swift-UVOT is related to the prevalence of the red giant emission in optical/near-IR wavelengths (see discussion in \citealt{Merc2023}). 
Similarly, \cite{Zamanov_2021} searched for rapid variability in SU~Lyn detecting flickering with a significant amplitude in the U band while no variability was detected in the B and V bands. 

\begin{figure*}
\centering
\includegraphics[scale=0.5]{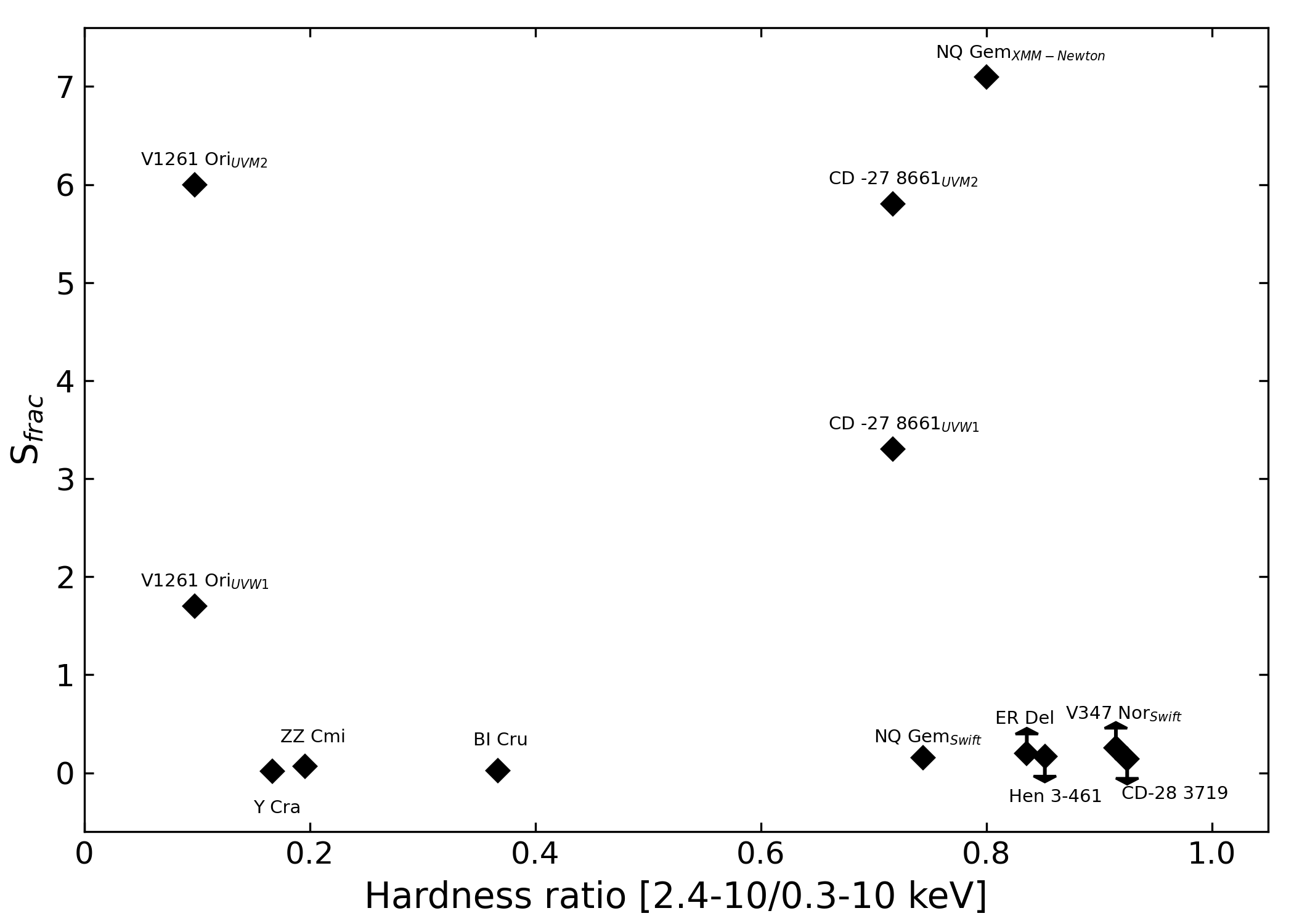}
\includegraphics[scale=0.5]{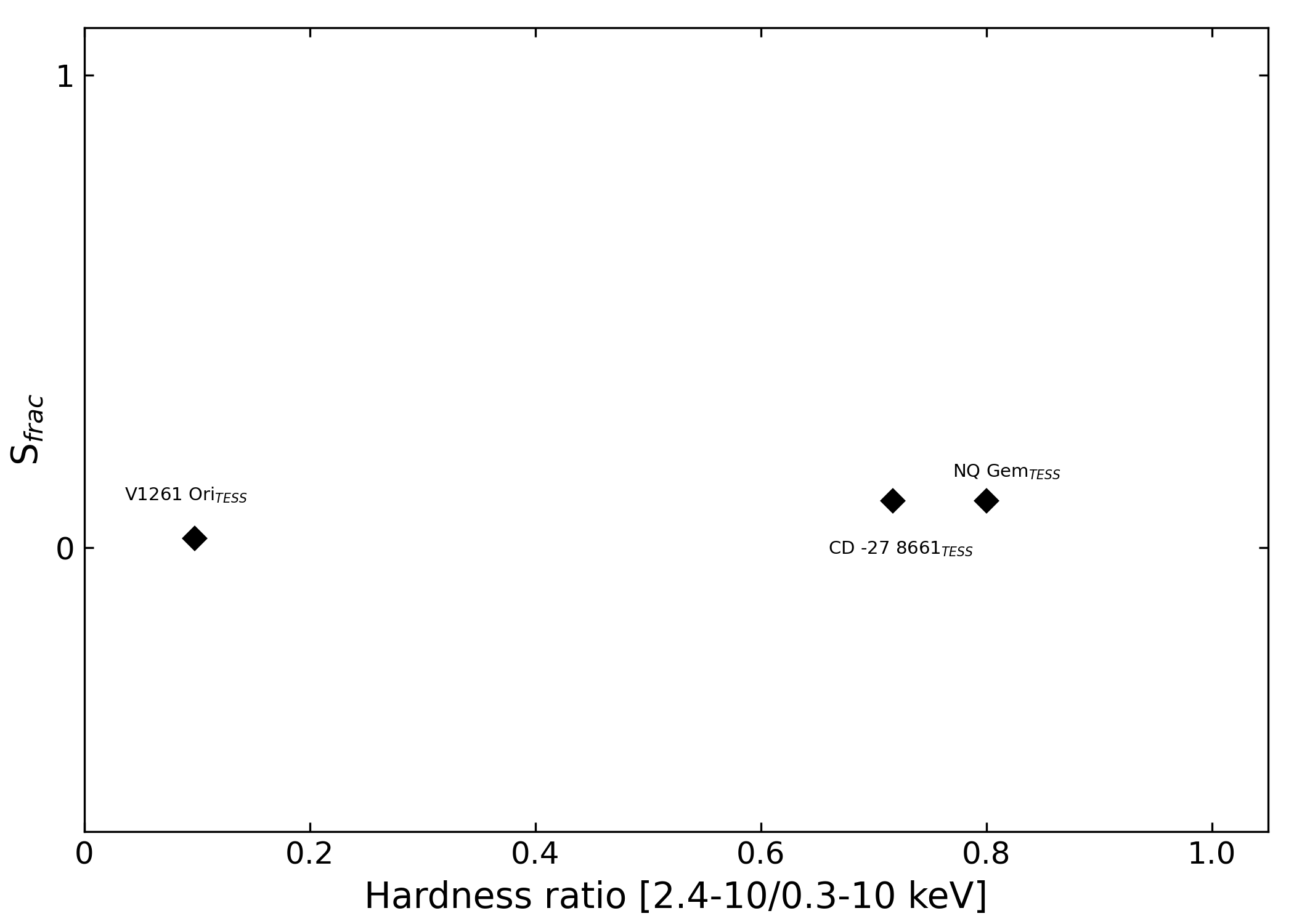}\\
\caption{Fractional amplitude of optical variability (S$_{frac}$) vs ratio of hard (2.4-10.0 keV) to total (0.3-10.0 keV) X-ray count rates from \cite{Luna_2013} and this paper.}
\label{fig:flick}
\end{figure*}

We found, in agreement with \cite{Luna_2013}, that symbiotics with harder X-ray emission, which are accretion-powered symbiotic stars, tend to exhibit greater amplitude of flickering, notably using bluer bands. Figure~\ref{fig:flick} shows an updated version of Fig.~6 from \cite{Luna_2013}, where we have included our measurements of the amplitude of flickering from UV and \tess\ data. Also, there is a clear lack of systems of intermediate and hard X-ray energies between 0.2 and 0.7~keV, which additional X-ray observations are needed for symbiotic stars.

\begin{table*}
\caption{Timing analysis from optical light curves given by each \tess\ section (before and after downlink intervals) and \xmm\ filter. $s$ is the measured standard deviation, $s_{exp}$ is the expected standard deviation from Poisson statistics, $s/s_{exp}$ is the ratio of standard deviations, and $s_{frac}$ is the percentage of the fractional variability amplitude.}
\centering
\label{tab:flickering}
\begin{tabular}{cccccccc} 
\hline\hline  
\noalign{\smallskip}
Objects  & Satellite & Section/Filter & s & s$_{exp}$ & s/s$_{exp}$  & s$_{frac}$ \\
& & & (mag) & (mag) & &  (\%)\\
\noalign{\smallskip}
\hline
\noalign{\smallskip}
V1261 Ori & \tess\  & 05 &  0.002 & 0.0001 & 2.7 & 0.02\\
           &      & 05 &  0.001 & 0.0001 & 3.1 & 0.02\\
           &      & 32 &  0.002 & 0.0001 & 2.1 & 0.02\\
           &      & 32 &  0.002 & 0.0001 & 2.1 & 0.02\\          
\\
       & \xmm\ & UVW1 & 0.017 & 0.04 & 0.5 & 1.7 \\
       & \xmm\ & UVM2 & 0.061 & 0.04 & 1.4 & 6.0 \\
\hline \noalign{\smallskip} 
NQ Gem & \tess\ & 44 & 0.001 & 0.0001 & 9.0 & 0.1\\ 
       &      & 44 & 0.001 & 0.0001 & 8.9 & 0.1\\ 
       &      & 45 & 0.001 & 0.0001 & 8.9 & 0.1\\
       &      & 45 & 0.001 & 0.0001 & 8.1 & 0.1\\
       &      & 46 & 0.001 & 0.0001 & 11.1 & 0.1 \\
       &      & 46 & 0.001 & 0.0001 & 9.3 & 0.1\\  
       &      & 71 & 0.0001 & 0.0003 & 0.3 & 0.01 \\
       &      & 71 & 0.0001 & 0.0001 & 0.3 & 0.01\\  
       &      & 72 & 0.0001 & 0.0003 & 0.3 & 0.01\\        
       &      & 72 & 0.0001 & 0.0003 & 0.3 & 0.01\\          
\\       
       & \xmm\ & UVW1 & 0.07 & 0.04 & 2.0 & 7.0 \\
       & \xmm\ & UVM2 & 0.07 & 0.03 & 2.6 & 7.2 \\
\hline \noalign{\smallskip} 
CD~-27~8661 & \tess\ & 10 & 0.001 & 0.0001 & 23.8 & 0.2\\ 
       &      & 10 & 0.001 & 0.0001 & 16.8 & 0.1\\ 
       &      & 37 & 0.001 & 0.0003 & 4.4 & 0.1\\ 
       &      & 37 & 0.001 & 0.0003 & 4.4 & 0.1\\ 
       &      & 64 & 0.001 & 0.0002 & 5.2 & 0.1\\ 
\\
       & \xmm\ & UVW1 & 0.03 & 0.04 & 0.9 & 3.0 \\
       & \xmm\ & UVW1 & 0.04 & 0.03 & 1.1 & 3.7\\
       & \xmm\ & UVW1 & 0.03 & 0.03 & 1.0 & 3.1 \\     
       & \xmm\ & UVM2 & 0.07 & 0.03 & 1.1 & 7.6 \\
       & \xmm\ & UVM2 & 0.03 & 0.03 & 1.0 & 2.6 \\     
       & \xmm\ & UVM2 & 0.07 & 0.03 & 2.6 & 7.3 \\
       & \xmm\ & UVM2 & 0.05 & 0.03 & 2.0 & 5.5\\      
\hline \noalign{\smallskip}
\end{tabular} 
\end{table*}

\section{Conclusions} \label{sec:con}

We present our study of four symbiotic stars: BD~Cam, V1261~Ori, NQ~Gem, and CD~-27~8661 using observations from the \swift-XRT and \xmm\ in X-rays and UV and \tess\ in optical.
Those symbiotic stars belong to the accretion-powered type, a finding supported by their hard X-ray emission as well as the presence of flickering in their light curves from \tess\ data.
Indeed, the $\delta$-type X-ray emission is present in those symbiotics that are powered by accretion, arising in the boundary layer. On the other hand, the soft component, $\beta$-type, originates from shocks, possibly between the red giant and WD/disk winds. And, the $\beta$/$\delta$-type are composed by soft and hard components, indicating that some $\beta$-type emission can also be produced by accretion-powered symbiotic stars. 
\tess\ data can be a useful diagnostic tool to search for flickering, which we have found to be a proxy for the hardness of the X-ray emission and the identification of accretion-powered symbiotic stars. 

\begin{acknowledgements}
The authors thank the referee for corrections and suggestions to the manuscript. IJL and GJML acknowledge support from grant ANPCYT-PICT 0901/2017. GJML and NEN are members of the CIC-CONICET (Argentina).
ASO acknowledges São Paulo Research Foundation (FAPESP) for financial support under grant \#2017/20309-7. JLS acknowledges support from NASA award 80NSSC21K0715. NP thanks the Coordenação de Aperfeiçoamento de Pessoal de Nível Superior -- Brazil (CAPES) for the financial support under grant 88887.823264/2023-00.
We thank Dr. Kim L. Page for help with the optical loading. 
We acknowledge the use of public data from the \swift\,  \xmm\, and \tess\ data archive.
We use of the \textit{python} packages \textit{matplotlib}, \textit{scipy}, \textit{numpy}, \textit{pyXSPEC}, and Bayesian X-ray Analysis (BXA).
We also acknowledge the variable star observations from the {\sc AAVSO} International Database contributed by observers worldwide and used in this research, the {\sc ASAS-SN} Sky Patrol, the {\sc DASCH} project at Harvard, which is partially supported by NSF grants AST-0407380, AST-0909073, and AST-1313370.

\end{acknowledgements}

%
\bibliography{ref} 
%


\begin{appendix} 

\section{X-ray spectral classifications of symbiotic stars} \label{appendix1}

\begin{table*}
\caption[]{X-ray spectral classifications of symbiotic stars.}
\centering
\label{tab:x_ray_list}
\begin{tabular}{cccccc} 
\hline\hline  
\noalign{\smallskip}
Objects  & Type & Reference \\
\noalign{\smallskip}
\hline
\noalign{\smallskip}
StHA 32 & $\alpha$  & \cite{Orio_2007, Luna_2013} \\ 
SMC 3 & $\alpha$ & \cite{Muerset_1997}\\ 
Lin 358  & $\alpha$ & \cite{Muerset_1997}\\ 
AG Dra  & $\alpha$ & \cite{Muerset_1997}\\ 
Draco C-1  & $\alpha$ & \cite{Muerset_1997, Saeedi_2018} \\ 
RR Tel  & $\alpha$ & \cite{Muerset_1997}\\ 
CD -43 14304  & $\alpha$ & \cite{Muerset_1997}\\ 
LMC S63 &  $\alpha$ & \cite{Bickert_1993}\\
\hline\\
BD Cam & $\beta^{\ast}$ & This work \\
SWIFT J171951.7-300206  & $\beta$ & \cite{Luna_2013} \\ 
RX Pup  & $\beta$ & \cite{Muerset_1997,Luna_2006}\\ 
Z And  & $\beta$  & \cite{Muerset_1997, Sokoloski_2006}\\ 
V1329 Cyg  &  $\beta$ & \cite{Stute_2011}  \\ 
HM Sge  & $\beta$ & \cite{Muerset_1997}\\ 
V1016 Cyg  & $\beta$ & \cite{Muerset_1997}\\ 
PU Vul  & $\beta$  & \cite{Muerset_1997}\\ 
AG Peg  & $\beta$  & \cite{Muerset_1997}\\ 
Hen 3-1341  & $\beta$ & \cite{Stute_2013}\\ 
Hen 2-87 & $\beta$ & \cite{nunez_2014}\\ 
NSV 25735 & $\beta$ & \cite{nunez_2014}\\ 
V2416 Sgr & $\beta$ & \cite{nunez_2014}\\ 
$o$ Cet & $\beta$ & \cite{Sokoloski_2010}\\
V1534 Sco & $\beta$ & \cite{Kuulkers_2014}\\
V745 Sco & $\beta$ & \cite{Page_2015,Merc_2021}\\
V852 Cen  & $\beta$ & \cite{Montez_2006,nunez_2014} \\
AS 201 & $\beta$ & \cite{Allen_1981,Bickert_1996}\\
2SXPS J173508.4-292958 &  $\beta$ & \cite{Munari_2020}\\
V426 Sge &  $\beta$ & \cite{Parikh_2018}\\
V2428 Cyg &  $\beta$ & \cite{Luna_2015a}\\
V407 Cyg  &  $\beta$ & \cite{Mukai_2012}\\
\hline\\
GX 1+4 & $\gamma$ & \cite{Muerset_1997}\\
V934 Her & $\gamma$ & \cite{Masetti_2002}\\ 
Sct X-1 & $\gamma$  & \cite{Kaplan_2007}\\ 
IGR J16194-2810 & $\gamma$ & \cite{Masetti_2007}\\ 
IGR J17329-2731 & $\gamma$ & \cite{Bozzo_2018} \\ 
2XMM J174445.4-295046 & $\gamma$ & \cite{Bahramian_2014}\\ 
SRGA J181414.6-225604 & $\gamma$ & \cite{2022ApJ...935...36D}\\
\hline\\
V1261 Ori  & $\delta$ & \cite{Jorissen_1996, Merc_2021}, This work\\
CD -27 8661 & $\delta$ & This work\\
ER Del  & $\delta$ & \cite{Luna_2013}\\ 
CD -28 3719  & $\delta$ & \cite{Luna_2013, nunez_2016}\\ %
RT Cru  & $\delta$ & \cite{Luna_2007,Kennea_2009} \\ 
T CrB &  $\delta$ & \cite{Luna_2008}\\ 
V648 Car & $\delta$ & \cite{Smith_2008, Eze_2010}\\ 
Hen 3-1591 & $\delta$ & \cite{nunez_2016}\\ 
4 Dra  & $\delta$ & \cite{nunez_2016}\\ 
EG And  &  $\delta$ & \cite{nunez_2016}\\ 
SU Lyn & $\delta$ & \cite{Mukai_2016}\\
ASAS J152058-4519.7 & $\delta$ & \cite{2021PhDT........17L}\\
Haro 1-10 & $\delta$ & \cite{2021PhDT........17L}\\
\hline\\
NQ Gem  & $\beta$/$\delta$ & \cite{Luna_2013,Toala_2023}, This work \\ 
BI Cru  & $\beta$/$\delta$ & \cite{Luna_2013}\\ 
UV Aur  & $\beta$/$\delta$ & \cite{Luna_2013}\\ 
ZZ CMi  & $\beta$/$\delta$  & \cite{Luna_2013}\\ 
V347 Nor  & $\beta$/$\delta$ & \cite{Luna_2013}\\
R Aqr  & $\beta$/$\delta$ & \cite{Muerset_1997,Nichols_2007}\\ 
CH Cyg  & $\beta$/$\delta$ & \cite{Muerset_1997,Mukai_2007} \\ 
V694 Mon  & $\beta$/$\delta$  & \cite{Stute_2009}\\ 
Hen 3-461  & $\beta$/$\delta$ & \cite{Luna_2013,nunez_2016} \\ 
\hline \noalign{\smallskip}
\end{tabular} \\
${\ast}$ BD Cam can be a $\delta$-type.
\end{table*}

\newpage

\section{X-ray Bayesian analyze} \label{appendix2}

\begin{figure*}
\begin{center}
\includegraphics[scale=0.4]{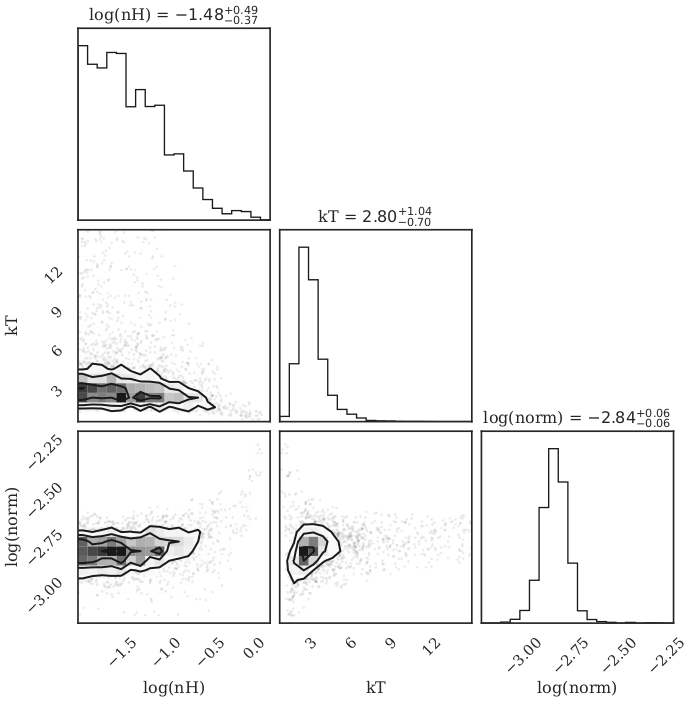}\includegraphics[scale=0.4]{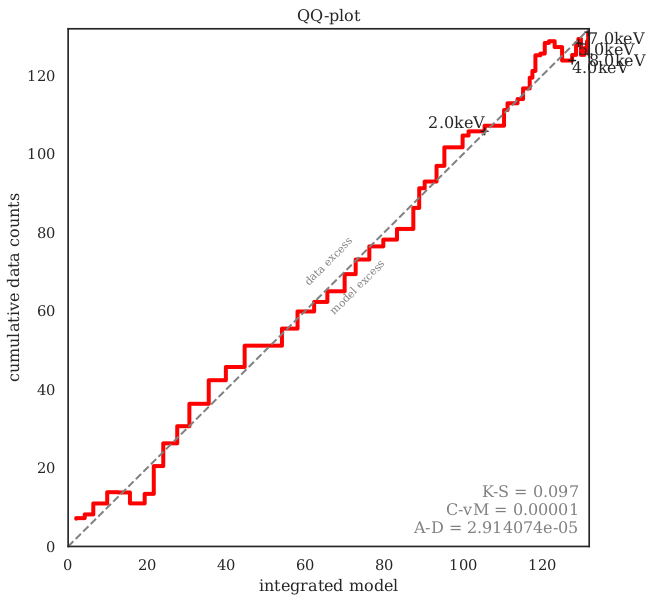}\\BD Cam\\
\includegraphics[scale=0.4]{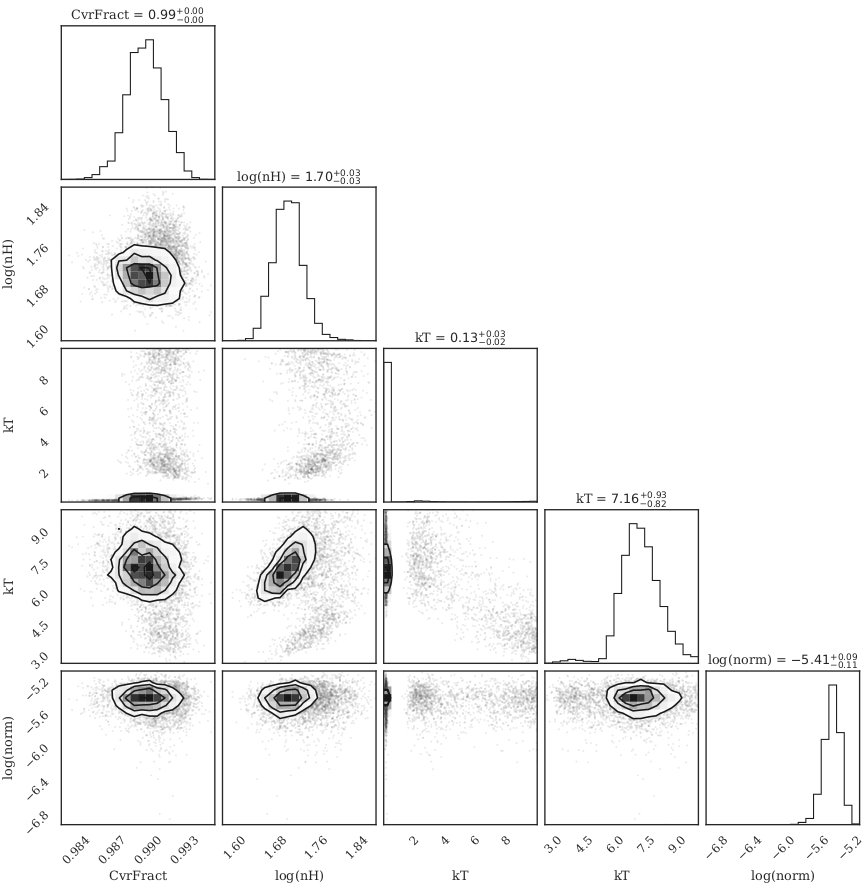}\includegraphics[scale=0.16]{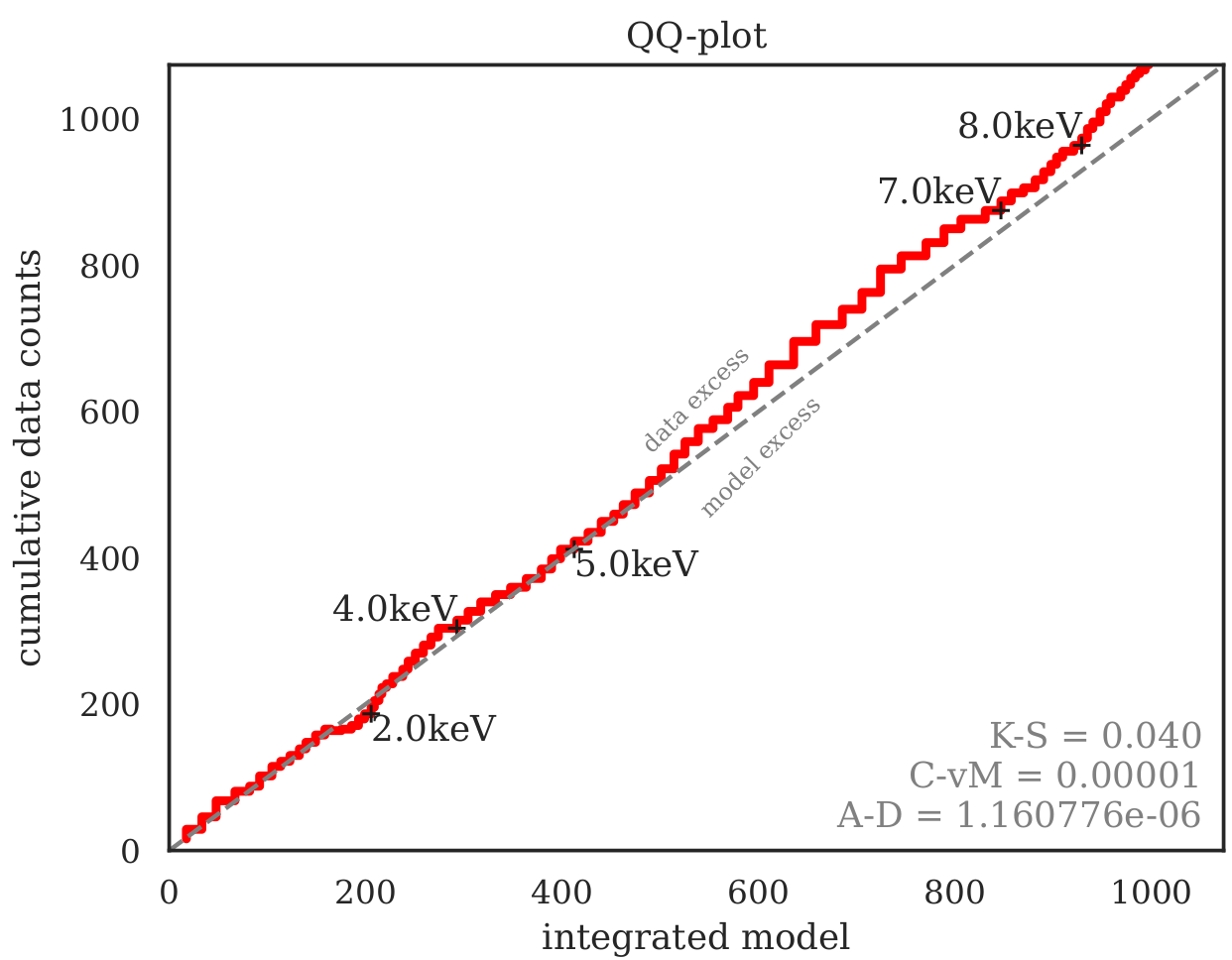}\\NQ Gem\\
\caption{Corner plots of the fitted parameters obtained from the Bayesian X-ray (BXA) algorithm. The probability densities are showed with shaded contoured region and its histogram indicating the average and error values for each parameters. The $Q~–~Q$ (quantile-quantile) plot are also present for each convergent results. Some statistical tests are also calculated Kolmogorov-Smirnov ($K-S$), Cramér–von Mises ($C-vM$), and the Anderson–Darling ($A-D$).}
\label{bxa_results_corner}
\end{center}
\end{figure*}

\begin{figure*}
\begin{center}
\includegraphics[scale=0.4]{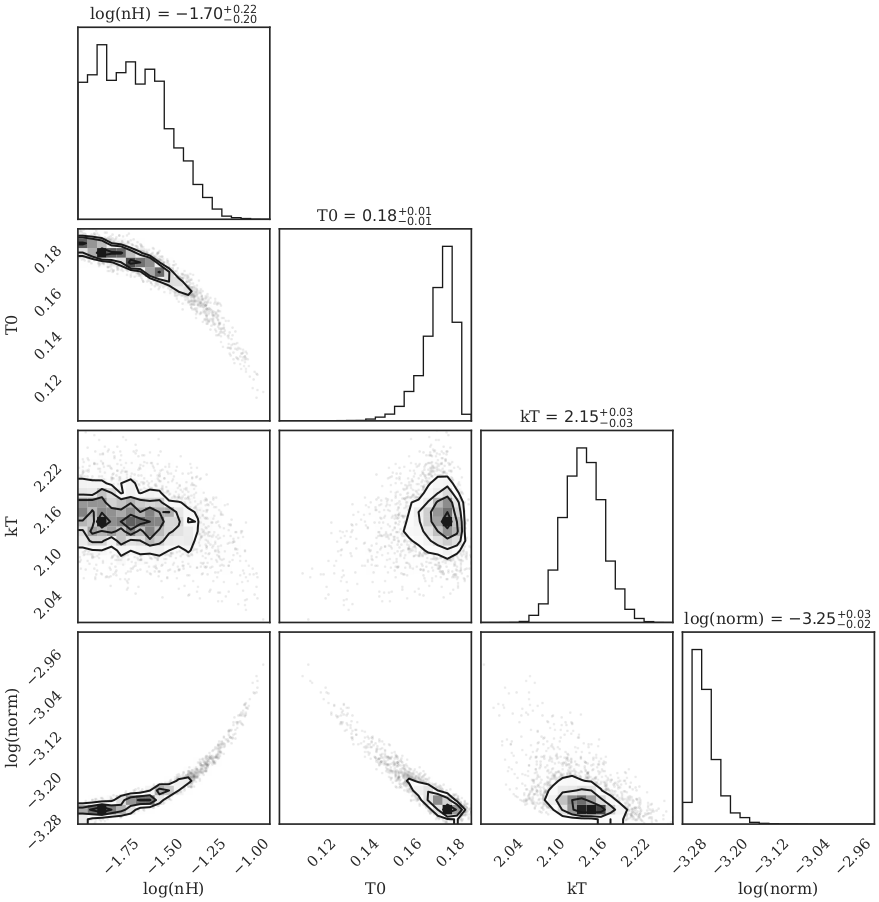}\includegraphics[scale=0.31]{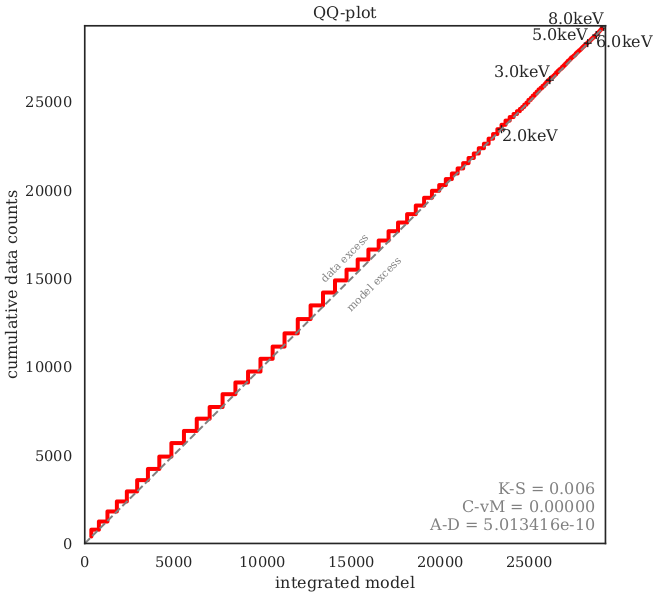}\\V1261 Ori: \texttt{tbabs~$\times$~compTT}\\
\includegraphics[scale=0.4]{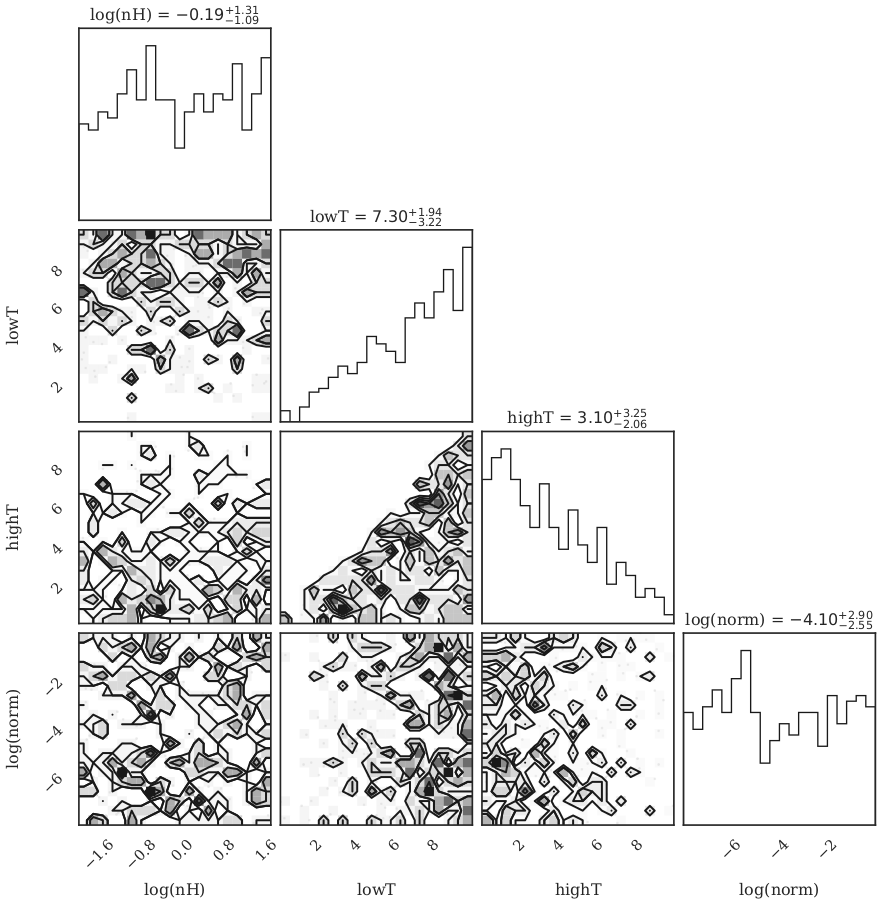}\\V1261 Ori: \texttt{tbabs~$\times$~mkcflow}\\
\caption{Same as Fig~\ref{bxa_results_corner}.}
\label{bxa_results_corner2}
\end{center}
\end{figure*}

\begin{figure*}
\begin{center}
\includegraphics[scale=0.4]{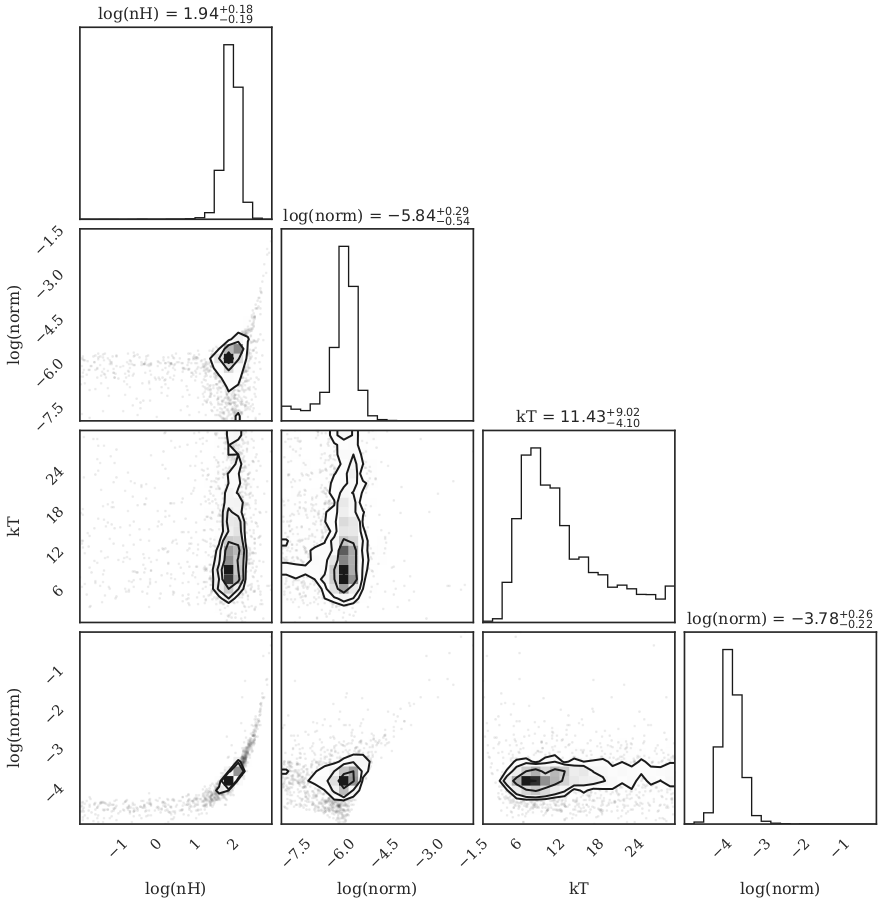}
\includegraphics[scale=0.4]{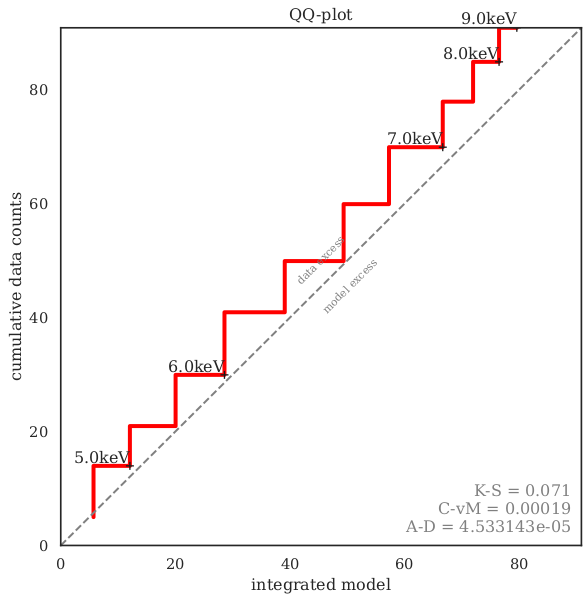}\\CD -27 8661\\
\caption{Same as Fig~\ref{bxa_results_corner}.}
\label{bxa_results_corner3}
\end{center}
\end{figure*}
\end{appendix}

\end{document}